\documentclass[prd,amsmath,amssymb,longbibliography,superscriptaddress,twocolumn,nofootinbib,10pt]{revtex4}

\pdfoutput=1

\usepackage{graphicx}
\usepackage{dcolumn}
\usepackage{bm}
\usepackage{amssymb}
\usepackage{latexsym}
\usepackage{booktabs}
\usepackage{amsmath}
\usepackage{multirow}
\usepackage[colorlinks,linkcolor=red,anchorcolor=red,citecolor=blue,hyperfootnotes=true]{hyperref}

\newcommand{\LCDM}{\rm{\Lambda}CDM}

\newcommand{\Mpc}{\mathrm{~km~s^{-1}~Mpc^{-1}}}

\begin{document}

\title{Torsion Cosmology in the Light of DESI, Supernovae and CMB Observational Constraints}

\author{Tonghua Liu}
\affiliation{School of Physics and Optoelectronic, Yangtze University, Jingzhou 434023, China;}
\author{Xiaolei Li}
\email{lixiaolei@hebtu.edu.cn}
\affiliation{College of Physics, Hebei Normal University, Shijiazhuang 519082, China; }
\author{Tengpeng Xu}
\affiliation{School of Physics and Astronomy, Sun Yat-sen University, Zhuhai 050024, China;}
\author{Marek Biesiada}
\affiliation{National Centre for Nuclear Research, Pasteura 7, PL-02-093 Warsaw, Poland; \\}
\author{Jieci Wang}
\email{jcwang@hunnu.edu.cn}
\affiliation{Department of Physics, and Collaborative Innovation Center for Quantum Effects and Applications, Hunan Normal University, Changsha 410081, China;}

\begin{abstract}
In this work, we investigate a torsion-based cosmological model within the Einstein–Cartan framework, constrained by the latest combined datasets including DESI DR2 BAO, PantheonPlus and DESY5 supernovae, and the full Planck 2018 CMB measurements (temperature, polarization, and joint NPIPE PR4 + ACT DR6 lensing). The torsion parameter is constrained to $\alpha = -0.00066 \pm 0.00098$ with the full dataset combination, consistent with zero at less than $1\sigma$, while yielding a Hubble constant $H_0 = 68.41 \pm 0.32~\mathrm{km~s^{-1}~Mpc^{-1}}$ and matter clustering amplitude $S_8 = 0.812 \pm 0.006$. The model shows notable potential in alleviating cosmological tensions, reducing the $S_8$ discrepancy with KiDS-1000 from $\sim 2.3\sigma$ in $\Lambda$CDM to only $0.1\sigma$. Model comparisons based on the Akaike information criterion show consistent improvements across all datasets, with $\Delta\mathrm{AIC}$ values ranging from $-5.68$ to $-6.62$, indicating a statistically preferred fit for the torsion model. These results suggest that the torsion framework provides a physically well-motivated extension to $\Lambda$CDM, capable of simultaneously addressing key cosmological tensions while maintaining excellent agreement with diverse observational probes.
\end{abstract}

\maketitle
\section{Introduction}
The discovery of cosmic acceleration through observations of Type Ia supernovae (SN Ia) \citep{1998AJ....116.1009R,1999ApJ...517..565P}  revolutionized our understanding of the universe's expansion, establishing the $\LCDM$ cosmological model as the prevailing framework. This model, validated through precise measurements of cosmic microwave background (CMB) anisotropies  \citep{2020A&A...641A...6P} and large-scale structure surveys \citep{2021PhRvD.103h3533A}, posits a cosmology dominated by cold dark matter and dark energy in the form of a cosmological constant. Nevertheless, increasingly precise measurements have revealed growing inconsistencies between different cosmological probes, with two particularly notable discrepancies emerging. First, the Hubble tension manifests as a statistically significant divergence between the expansion rate derived from early-universe CMB measurements ($H_0$ = 67.4 $\pm$ 0.5 $\Mpc$) \citep{2020A&A...641A...6P} and late-universe distance ladder determinations ($H_0$ = 73.04 $\pm$ 1.04$\Mpc$) by SH0ES team \citep{2022ApJ...934L...7R}. Second, the $\sigma_8$ tension, and associated $S_8$ tension reflect a disagreement in matter clustering measurements, with CMB-based estimates exceeding those from weak gravitational lensing surveys. These persistent anomalies have stimulated vigorous research into alternative cosmological scenarios, including dynamic dark energy formulations and modifications to general relativity. The scientific community has produced extensive literature examining these tensions, with particular attention to the Hubble constant discrepancy  \citep{2019PhRvL.122v1301P,2019PhRvL.122f1105F,
2023arXiv230306974D,2022ApJ...939...37L,2021ApJ...912..150D,2020ApJ...895L..29L,2023ApJS..264...46L,2020PhRvD.102b3518V,2023Univ....9..393V,2021PhRvD.103h1305B,2022JCAP...04..004L}. Various theoretical approaches have been proposed to reconcile these observational differences, ranging from early dark energy models to novel gravitational theories. Current investigations continue to explore whether these tensions indicate new physics beyond the standard cosmological model or stem from systematic effects in observations.

The ongoing five-year survey conducted by the Dark Energy Spectroscopic Instrument (DESI) has provided unprecedented insights into cosmological tensions through its high-precision measurements \citep{2025JCAP...02..021A}. Analysis of the second data release (DR2) \citep{2025arXiv250314738D}, incorporating baryon acoustic oscillation (BAO) signals from over 14 million extragalactic sources alongside complementary cosmological datasets, demonstrates compelling evidence for dynamical dark energy evolution. These results indicate a statistically significant departure ($2.8-4.2\sigma$) from the cosmological constant paradigm, favoring instead an equation of state parameter that evolves with redshift  ($w_0 > -1,  w_a < 0$). While these findings offer compelling evidence for dynamical dark energy, the CPL parameterization remains an empirical framework without fundamental physical motivation. This has prompted researchers to investigate more theoretically grounded explanations for the observed cosmic acceleration, including modified gravity theories and quantum field-based dark energy models \citep{2025PhRvL.134r1002Y,2025arXiv250524732C,2025PhRvD.111h3523F,2025arXiv250407791L,2024arXiv241112026I,2024arXiv241111743Y}.  Current investigations focus on developing self-consistent theoretical frameworks that can simultaneously explain the DESI observations while addressing other cosmological tensions. Recently, the Interacting Dark Energy (IDE) models provided results that were equally good or better than the standard $\LCDM$ model for both high- and low-redshift observations. However, these models face a key limitation: their predictions of lower matter density and higher matter clustering amplitude may conflict with large-scale structure measurements  \citep{2024PhRvL.133y1003G,2025PhRvD.111l3511S,2025arXiv250321652S,2024ApJ...976....1L}.
Similarly,  based on the new data of DESI, various alternative models and methodologies  that attempt to solve the  aforementioned issues have also been  extensively investigated in the literature. For a non-exhaustive set of references on this topic, see \citep{2025arXiv250400985Y,2025arXiv250321652S,2025arXiv250319898P,2025arXiv250319352I,2024ApJ...976....1L,2025arXiv250314743L,2025arXiv250421373L,2025arXiv250502932C,2024ApJ...976L..11R,2025ApJ...986L..31R,
2025arXiv250204212H,2025arXiv250407679W,2025arXiv250416868A,2025arXiv250310806C,2025arXiv250416868A,
2024PhRvD.110l3533P,2024arXiv241013627P,2025arXiv250103480P,2022PhRvD.106e5014Y,
2024JHEP...05..327Y,2025arXiv250318417L,2024PhRvD.110l3519J,2025JCAP...01..153J,2011PhLB..698..175C,2025arXiv250404646M,2025arXiv250621542G,
2025arXiv250404417C,2025arXiv250324343C,2025arXiv250219274W,2025arXiv250321600P,2025arXiv250322529N,2025arXiv250323225S,2025JCAP...01..153J}.

Inspired by the aforementioned studies, we propose a novel cosmological model within the framework of Einstein-Cartan (EC) theory. The model is grounded in a profound physical insight, i.e., the torsion effect of spacetime essentially represents a macroscopic manifestation of matter's intrinsic angular momentum (spin) \citep{2002RPPh...65..599H,1973GReGr...4..333H,1975PhDT.......104K}. This theoretical construction demonstrates several remarkable advantages.
First, it possesses a solid theoretical foundation, being entirely built upon the geometric framework of Einstein-Cartan theory without requiring any ad hoc assumptions or additional fields, while the spin-torsion coupling has a clear quantum field theoretical basis. Second, the model naturally incorporates dynamic dark energy behavior through the torsion term \citep{2019EPJC...79..341K}, offering a geometric mechanism for cosmic acceleration. In this work, we investigate its potential to simultaneously alleviate the Hubble tension and $S_8$ tension, and test its consistency with the latest observational data from DESI, PantheonPlus, and CMB surveys.

This paper is organized as follows:  Section \ref{sec:2} outlines the Einstein-Cartan field equations and the resulting Friedmann equations for a  homogeneous and isotropic Universe. In Section \ref{sec:3}, we present the observational data used in this work.  We give our results and discussion in Section \ref{sec:4}. Finally, the main conclusions are summarized in Section \ref{sec:5}.

\section{Homogeneous, isotropic Universe in the Einstein-Cartan-Kibble-Sciama cosmology}\label{sec:2}

The EC theory was based on a simple physical intuition, that torsion effect is regarded as a macroscopic manifestation of the intrinsic angular momentum (spin) of matter. \citet{1961JMP.....2..212K} and \citet{1964RvMP...36..463S} then reintroduced the spin of matter independently into GR in what is also known as ECKS theory. The ECKS theory postulates an asymmetric affine connection for the spacetime, in contrast to the symmetric Christoffel symbols of Riemannian spaces. In technical terms, torsion is described by the antisymmetric part of the non-Riemannian affine connection.
For convenience, we adopt the same notation as \citet{2019EPJC...79..341K}.

The Einstein-Cartan field equations with torsion are given by:
\begin{equation}
G_{\mu\nu} = \kappa T_{\mu\nu} - \Lambda g_{\mu\nu} + \kappa \tau_{\mu\nu},
\end{equation}
where $\kappa=8\pi G$ and the spin correction term is:
\begin{equation}
\tau_{\mu\nu} = -4 S_\mu S_\nu + 2 g_{\mu\nu} S_\alpha S^\alpha.
\end{equation} 

In  ECKS theory, the connection $\tilde{\Gamma}^\alpha_{\mu\nu}$ splits into the Levi-Civita part $\Gamma^\alpha_{\mu\nu}$ and the contortion tensor $K^\alpha_{\mu\nu}$:
\begin{equation}
\tilde{\Gamma}^\alpha_{\mu\nu} = \Gamma^\alpha_{\mu\nu} + K^\alpha_{\mu\nu},
\end{equation}
where the contortion tensor is given by:
\begin{equation}
K^\alpha_{\mu\nu} = S^\alpha_{\mu\nu} + S_{\mu\nu}^{\ \ \alpha} + S_{\nu\mu}^{\ \ \alpha}.\label{eq3}
\end{equation}
For a homogeneous, isotropic and flat universe, torsion tensor and torsion vector take the form:
\begin{align}
S_{\alpha\mu\nu} &= \phi(t)(h_{\alpha\mu}u_\nu - h_{\alpha\nu}u_\mu),\,\,
S_\alpha=-3\phi(t)u_\alpha, \label{eq4}
\end{align}
where $\phi(t)$ is a scalar function, $h_{\mu\nu} = g_{\mu\nu} + u_\mu u_\nu$ is the projection tensor, and  $u_\mu = (-1,0,0,0)$ is the 4-velocity.
Combing the Eqs. (\ref{eq3}) and (\ref{eq4}), we have:
\begin{equation}
K^{\alpha}{}_{\mu\nu} = \phi(t)(u^{\alpha}g_{\mu\nu} - u_{\mu}\delta^{\alpha}_{\nu}),
\end{equation}
and the non-zero contortion components are:
\begin{align}
K^{0}_{\ ij}=2\phi(t) a^2 \delta_{ij}, \,\,
K^{i}_{\ 0j}=2\phi(t) \delta^{i}_{j}.
\end{align}

Assuming a perfect fluid with the energy-momentum tensor $T_{\mu\nu}=\rho u_{\mu}u_{\nu}+ph_{\mu\nu}$, one has:
\begin{equation}
H^2 = \frac{\kappa}{3}\rho - 4\phi(t)^2-4\phi(t)H, \label{Fried}
\end{equation}
where $\rho$ and $p$ are the energy density and pressure of the matter fluid, respectively, and $H=\dot{a}/a$ denotes the Hubble function (i.e. the expansion rate of the Universe), see the Refs. \citet{1961JMP.....2..212K,1964RvMP...36..463S,2019EPJC...79..341K}. When $\phi=0$, the standard Friedman equation 
is recovered. 
Moreover, in an empty, flat ECKS Universe without cosmological constant, the Eq. (\ref{Fried}) reduces to $(H+2 \phi)^2=0$ and implies that $\phi$ is proportional to the expansion rate $H$ and has a analytical solution $\phi(t)=-\frac{1}{2} H$. Therefore, we assume the following parametrization  $\phi(t)=-\frac{1}{2}\alpha H$. 
In dark energy (cosmological constant) dominated time, Friedman Equation (\ref{Fried}) reads:
\begin{equation}
H^2(z) = \frac{\kappa}{3}[\rho_m + \rho_\Lambda] - \alpha^2 H(z)^2+ 2\alpha H(z)^2.
\end{equation}

In the standard ECKS theory, dark energy is not directly coupled to the torsion.
The conservation equation obtained by combining the  ECKS field equations and the twice-contracted Bianchi identities in  ECKS cosmology takes the following form:
\begin{equation}
\nabla_\mu T^{\mu\nu} + K^\lambda_{\mu\lambda}T^{\mu\nu} - K^\nu_{\mu\lambda}T^{\mu\lambda} = 0.
\end{equation}

For the barotropic fluid with $p=0$ (pressure-less matter),  the torsion-modified continuity equation is:
\begin{equation}
\dot{\rho}_m + 3H\rho_m -\alpha H\rho_m=0,
\end{equation}
Solving this equation leads us to
\begin{equation}
\rho_m(a) = \rho_{m0} a^{-3+\alpha}.
\end{equation}
By defining the present matter density parameter $\Omega_{m}=\frac{\kappa\rho_{m0}}{3H_0^2}$ and an analogous parameter for the cosmological constant $\Omega_{\Lambda}=\frac{\Lambda}{3H_0^2}$ and using $a=1/(1+z)$, we have:
\begin{equation}
{H(z)}^2 = \frac{H_0^2}{(1-\alpha)^2}[\Omega_m(1+z)^{3-\alpha}+ \Omega_\Lambda].
\end{equation}
Friedman equation at $z=0$, gives $\Omega_{\Lambda}=1-2\alpha+\alpha^2-\Omega_m$ and in the limit of $\alpha\to0$ the standard $\Lambda$CDM model is recovered.


\section{ Data and Methodology}\label{sec:3}

The datasets used in the analyses are described below:

$\bullet$ BAO:
Our cosmological analysis utilizes the most recent BAO data from the DESI DR2 release, which represents the most comprehensive BAO measurement to date by combining multiple tracers: luminous red galaxies, emission line galaxies, quasars, and Lyman-$\alpha$ forest absorption systems. These measurements provide precise determinations of three key cosmological distance ratios: the transverse comoving distance $D_M/r_d$ (where $r_d$ is the sound horizon at the drag epoch), the Hubble distance $D_H/r_d$, and the volume-averaged distance $D_V/r_d$ - reported in Table IV of the second release (DR2) \citep{2025arXiv250314738D}.
To ensure rigorous statistical treatment, our methodology fully accounts for the complete covariance structure of these measurements, including the significant cross-correlation coefficient $r_{M,H}$  between the transverse and line-of-sight distance measurements (typically ranging from 0.3 to 0.5 for different redshift bins). We denote this full dataset as \textbf{``DESI"}.

$\bullet$ SN Ia: We adopt SN Ia data from two compilations, Pantheon Plus and Dark Energy Survey (DES) 5 year data.
Our analysis incorporates the full Pantheon Plus  SN Ia sample \citep{2022ApJ...938..113S}, comprising 1,701 high-quality light curves from 1,550 spectroscopically confirmed  SN Ia  spanning an extensive redshift range ($0.01<z<2.26$). To minimize potential systematics from calibration uncertainties at low redshifts, we implement two key methodological choices: (1) we exclude the calibration subsample $(z < 0.01)$ that may be affected by peculiar velocity corrections and host galaxy contamination, and (2) we rigorously account for the full covariance matrix$\footnote{\url{https://github.com/PantheonPlusSH0ES/DataRelease}}$ that captures both statistical uncertainties and systematic correlations between supernova measurements. This conservative approach ensures our cosmological constraints remain robust against calibration-related biases while maintaining the statistical power of the full dataset. The remaining sample of 1,485 SN Ia after this selection provides a well-characterized Hubble diagram for precision cosmology.  On the other hand, the Dark Energy Survey (DES) collaboration has recently released a partial dataset from its full 5-year survey \citep{2024ApJ...973L..14D}, featuring a newly constructed, homogeneously selected sample of 1635 photometrically classified SN spanning the redshift range $0.1<z<1.3$. This sample is further augmented by an additional 194 low-redshift SN, resulting in a combined catalog of 1829 SN Ia $\footnote{\url{https://github.com/des-science/DES-SN5YR}}$. This release represents a significant step forward in the study of cosmic acceleration and dark energy, providing a more uniform and extensive dataset for cosmological analyses. We label these SN Ia  datasets as \textbf{``PantheonPlus"} and  \textbf{``DESY5"}.

\begin{table*}[htbp!]
\renewcommand\arraystretch{1.4}
  \centering
   \caption{Cosmological parameter constraints results with $1\sigma$ uncertainty  by using the CMB, CMB+DESI, and CMB+DESI+PantheonPlus/DESY5 datasets in $\Lambda$CDM. We also give the reduced minimum $\chi^2_{\mathrm{min}}/N_{dof}$, where $N_{dof}$ is the difference obtained by subtracting the number of model parameters from the total number of data points used. }
   \vspace{0.1cm}
  \begin{tabular}{c|cccc}
    \toprule
    \hline\hline
    \textbf{Model} & \multicolumn{4}{|c}{$\Lambda$CDM model} \\
    \midrule
    \hline
    \textbf{Data} & CMB & CMB+DESI & CMB+DESI+DESY5 & CMB+DESI+PantheonPlus \\
    \midrule
    \hline
    $H_0$ & $67.31\pm 0.53$ & $68.40\pm 0.29$ & $68.20\pm 0.29$ & $68.31\pm 0.29$\\
    \hline
$\Omega_\mathrm{b} h^2$ & $0.02238\pm 0.00015$ & $0.02256\pm 0.00013$ & $0.02252\pm 0.00012$ & $0.02254\pm 0.00013$\\
\hline
$\Omega_\mathrm{c} h^2$ & $0.1201\pm 0.0012$ & $0.11773\pm 0.00064$ & $0.11816\pm 0.00063$ & $0.11793\pm 0.00063$\\
\hline
$\log(10^{10} A_\mathrm{s})$ & $3.047\pm 0.013$ & $3.059^{+0.013}_{-0.014}$ & $3.057\pm 0.013$ & $3.058\pm 0.013$\\
\hline
$n_\mathrm{s}$ & $0.9654\pm 0.0041$ & $0.9714\pm 0.0034$ & $0.9703\pm 0.0033$ & $0.9709\pm 0.0033$\\
\hline
$\tau_\mathrm{reio}$ & $0.0551\pm 0.0075$ & $0.0625^{+0.0069}_{-0.0078}$ & $0.0610\pm 0.0073$ & $0.0617\pm 0.0075$\\
\hline
    \midrule
$\Omega_\mathrm{m}$ & $0.3160\pm 0.0073$ & $0.3013\pm 0.0037$ & $0.3038\pm 0.0036$ & $0.3025\pm 0.0036$\\
\hline
$\sigma_8$ & $0.8127\pm 0.0052$ & $0.8110\pm 0.0055$ & $0.8112\pm 0.0054$ & $0.8110\pm 0.0055$\\
    \midrule
\hline
$\chi^2_{\mathrm{min}}/N_{\rm dof}$ & $1.207$ & $1.208$ & $1.108$ & $1.016$ \\
\hline\hline
    \bottomrule
    \end{tabular}\label{tab1}
\end{table*}

$\bullet$ CMB: In our analysis, we employ the full set of \textit{Planck} 2018 likelihoods \citep{2020A&A...641A...6P,2020A&A...641A...5P,2020A&A...641A...1P}, which include the high-$\ell$ temperature and polarization spectra (high-$\ell$ TT+TE+EE), together with the low-$\ell$ temperature (low-$\ell$ TT) and polarization (low-$\ell$ EE) measurements. These datasets deliver precise determinations of the acoustic peak structure in the primary CMB anisotropies and are highly sensitive to key cosmological parameters such as the baryon density, matter content, and expansion history of the universe.
In addition to the primary anisotropy spectra, we also include measurements of the CMB lensing potential power spectrum $C_\ell^{\phi\phi}$. This signal arises from the four-point correlations of the observed CMB and traces the integrated distribution of large-scale structure along the line of sight, thereby probing late-time gravitational potentials and growth of structure \citep{Planck2013lensing,Planck2015lensing,Planck2018lensing}. For this work, we utilize the lensing likelihood constructed from the joint analysis of the \textit{Planck} PR4 (NPIPE) lensing reconstruction \citep{PlanckPR4} and the recent DR6 results from the Atacama Cosmology Telescope (ACT) \citep{ACTDR6lensing1,ACTDR6lensing2,ACTDR6lensing3}. We refer to this combined dataset as \textit{Planck}+ACT lensing.
Throughout this paper, \textbf{``CMB"} denotes the combination of \textit{Planck} 2018 TT, TE, and EE spectra (including both high- and low-$\ell$ likelihoods) together with the \textit{Planck}+ACT lensing likelihood. This comprehensive dataset constrains both the early-universe physics encoded in the primordial anisotropies and the late-time evolution probed by lensing, and thus provides a solid foundation for testing cosmological models.

This study incorporates multiple cosmological observational datasets. The specific numbers of data points are as follows: for the CMB data, these include the Planck high-$\ell$ TT+TE+EE spectra (2289 data points), the Planck low-$\ell$ TT spectra (27 data points), the Planck low-$\ell$ EE spectra (27 data points), and the Planck+ACT lensing potential power spectrum (27 data points). The BAO data consist of measurements from DESI R2 (12 data points) and DESY5 (1829 data points). Additionally, the PantheonPlus  SN Ia sample (1701 data points) is used.

We implement the theoretical model in a modified version of the \textsf{CLASS} Boltzmann solver code \citep{2011JCAP...07..034B} and employ the publicly available \textsf{Cobaya}$\footnote{\url{https://github.com/CobayaSampler/cobaya}}$ sampling framework \citep{2021JCAP...05..057T} for Bayesian parameter estimation through Markov Chain Monte Carlo (MCMC) analysis. In our analysis, we incorporate this code and combine it with the likelihoods from DESI BAO and SN Ia to study our target cosmological models.
To ensure the reliability of MCMC sampling, we implement the Gelman-Rubin convergence diagnostic criterion \cite{1992StaSc...7..457G}, enforcing a strict convergence threshold ($R-1<0.01$) across all models and datasets. For parameter estimation, we adopt uniform priors on the cosmological parameter set
$\{\Omega_bh^2, \Omega_ch^2,\tau_{reio},\theta_s,\log(10^{10}A_s),n_s,\alpha\}$ 
and perform statistical analysis of posterior distributions using the \textsf{GetDist} package$\footnote{\url{https://github.com/cmbant/getdist}}$, obtaining key statistical measures including one-dimensional marginalized distributions and two-dimensional joint confidence regions. 

Then, we compare the performance of our  torsion (ECKS) model against the standard $\Lambda$CDM cosmology by assessing the relative support from the observational data. A widely used criterion for such model comparison is the Akaike Information Criterion (AIC) \cite{2007JCAP...02..003B,2025PDU....4901965D}. The AIC is derived from an approximate minimization of the Kullback–Leibler information entropy, which quantifies the discrepancy between the true distribution of the data and the model distribution. It is defined as $AIC = - 2 \ln{{\cal L}_{max}} + 2 k$ where ${\cal L}_{max}$, where $\mathcal{L}_{\mathrm{max}}$ is the maximum likelihood attainable by the model and $k$ is the number of free parameters. The likelihood is commonly approximated as $\mathcal{L} \propto \exp(-\chi^2/2)$. Since the true underlying model is unknown, the absolute AIC value for a single model is not interpretable. Instead, the difference $\Delta\mathrm{AIC} = \mathrm{AIC} - \mathrm{AIC}_{\mathrm{baseline}}$ is used, where we take $\Lambda$CDM as the baseline. A lower AIC value indicates a preferred model: $\Delta\mathrm{AIC} > 0$ favors the baseline, while $\Delta\mathrm{AIC} < 0$ favors the test model (ECKS). The magnitude of $|\Delta\mathrm{AIC}|$ indicates the strength of the preference: $|\Delta\mathrm{AIC}| \geq 2$ is considered weak evidence, $|\Delta\mathrm{AIC}| \geq 6$ moderate evidence, and $|\Delta\mathrm{AIC}| \geq 10$ strong evidence \cite{2025PDU....4901965D}.
Another common metric, the Bayesian Information Criterion (BIC), was introduced by \cite{1978AnSta...6..461S} as an approximation to the Bayes factor. However, the BIC assumes that data points are independent and identically distributed—an assumption violated by correlated CMB data. For this reason, we do not employ the BIC in our analysis.

We assess the relative evidence for our torsion model against $\LCDM$ using the AIC. The $\LCDM$ model has 6 free parameters, while the torsion model extends it with one additional free parameter, $\alpha$, giving 7 free parameters in total. It is important to distinguish these fitted parameters from derived parameters such as $H_0$ and $\sigma_8$, whose values are consequentially determined by the fit but do not contribute to the model's degrees of freedom for the purpose of information criteria.

\section{Results and Discussion}\label{sec:4}

\begin{figure}
\begin{center}
{\includegraphics[width=0.95\linewidth]{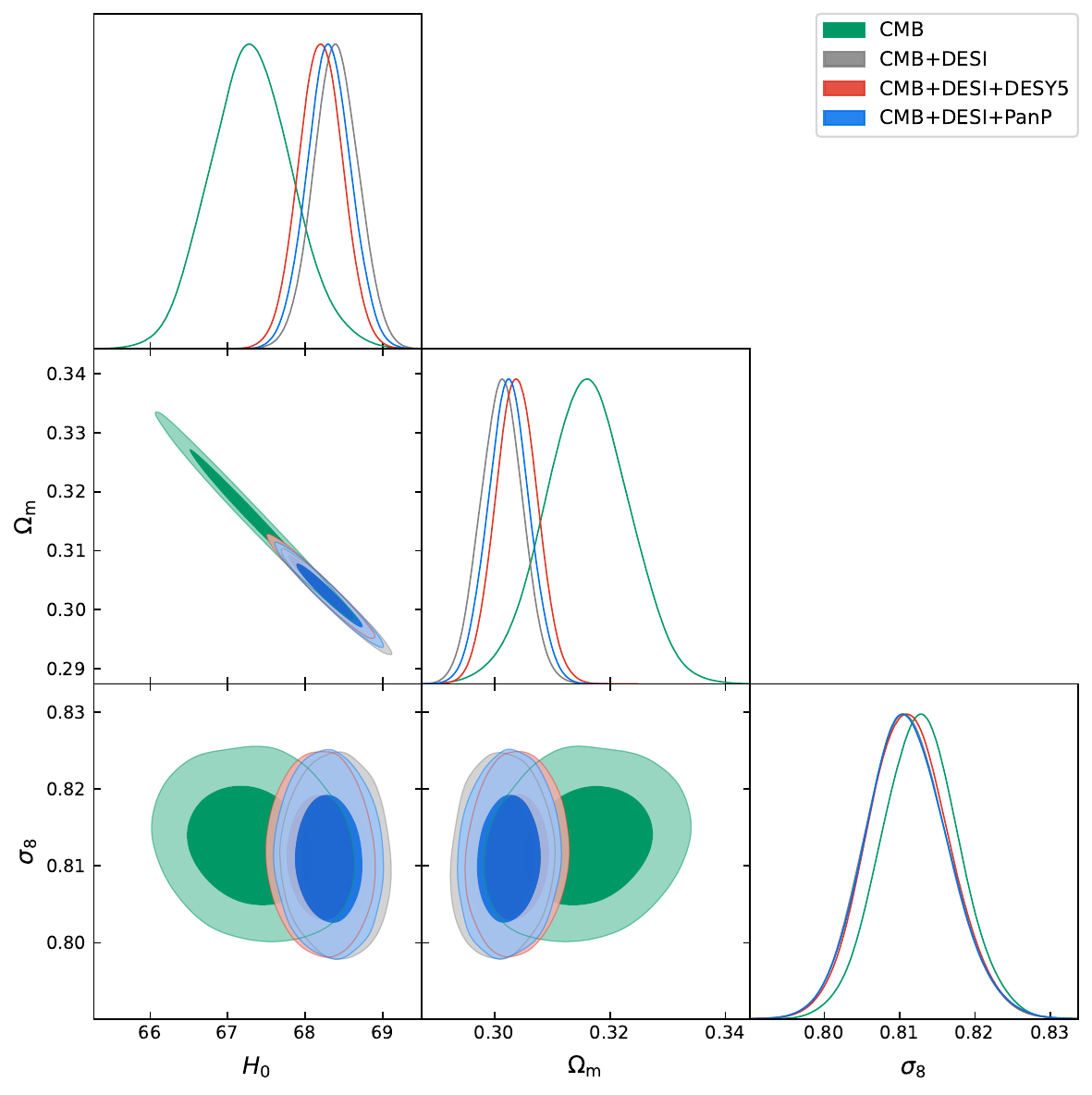}}
\end{center}
\caption{The one and two-dimensional marginalised posterior distributions in a $\Lambda$CDM cosmology from the different datasets.}  \label{torsion1}
\end{figure}

\begin{table*}[htbp!]
\renewcommand\arraystretch{1.4}
  \centering
  \caption{Cosmological parameter constraints results with $1\sigma$ uncertainty  by using the CMB, CMB+DESI, and CMB+DESI+PantheonPlus/DESY5 datasets. 
  Negative values of $\Delta$AIC=AIC$_{\rm torsion}$-AIC$_{\rm \LCDM}$ indicate a better fit and a preference for the torsion model over the $\LCDM$.}
\vspace{0.1cm}
  \begin{tabular}{c|cccc}
  \hline
  \hline
    \toprule
    \textbf{Model} & \multicolumn{4}{|c}{Torsion model} \\
    \midrule
    \hline
    \textbf{Data} & CMB & CMB+DESI & CMB+DESI+DESY5 & CMB+DESI+PantheonPlus \\
    \midrule
    \hline
    $\alpha$ & $0.0006\pm 0.0011$ & $-0.00083\pm 0.00099$ & $-0.00050\pm 0.00098$ & $-0.00066\pm 0.00098$\\
        \hline
$H_0$ & $67.06\pm 0.70$ & $68.54\pm 0.33$ & $68.27\pm 0.32$ & $68.41\pm 0.32$\\
\hline
$\Omega_\mathrm{b} h^2$ & $0.02243\pm 0.00017$ & $0.02245\pm 0.00018$ & $0.02245\pm 0.00017$ & $0.02245\pm 0.00018$\\
\hline
$\Omega_\mathrm{c} h^2$ & $0.1203\pm 0.0013$ & $0.11787\pm 0.00067$ & $0.11827\pm 0.00066$ & $0.11804\pm 0.00066$\\
\hline
$\log(10^{10} A_\mathrm{s})$ & $3.046\pm 0.013$ & $3.058^{+0.013}_{-0.014}$ & $3.056\pm 0.013$ & $3.057\pm 0.013$\\
\hline
$n_\mathrm{s}$ & $0.9640\pm 0.0048$ & $0.9723\pm 0.0035$ & $0.9708\pm 0.0034$ & $0.9716\pm 0.0034$\\
\hline
$\tau_\mathrm{reio}$ & $0.0541\pm 0.0074$ & $0.0623^{+0.0068}_{-0.0080}$ & $0.0608^{+0.0067}_{-0.0075}$ & $0.0614\pm 0.0073$\\
    \midrule
\hline  
$\Omega_\mathrm{m}$ & $0.3190\pm 0.0092$ & $0.3001\pm 0.0039$ & $0.3033\pm 0.0039$ & $0.3016\pm 0.0038$\\
\hline
$\sigma_8$ & $0.8115\pm 0.0054$ & $0.8126\pm 0.0059$ & $0.8123\pm 0.0057$ & $0.8123\pm 0.0057$\\
    \midrule
 \hline
$\chi^2_{\mathrm{min}}/N_{\rm dof}$ & $1.204$ & $1.205$ & $1.106$ & $1.013$\\
\hline
$\Delta$AIC  & $-6.16$ & $-5.68$ & $-6.04$ & $-6.62$\\
\hline
\hline
    \bottomrule
    \end{tabular}\label{tab2}
\end{table*}

\begin{figure}
\begin{center}
{\includegraphics[width=0.95\linewidth]{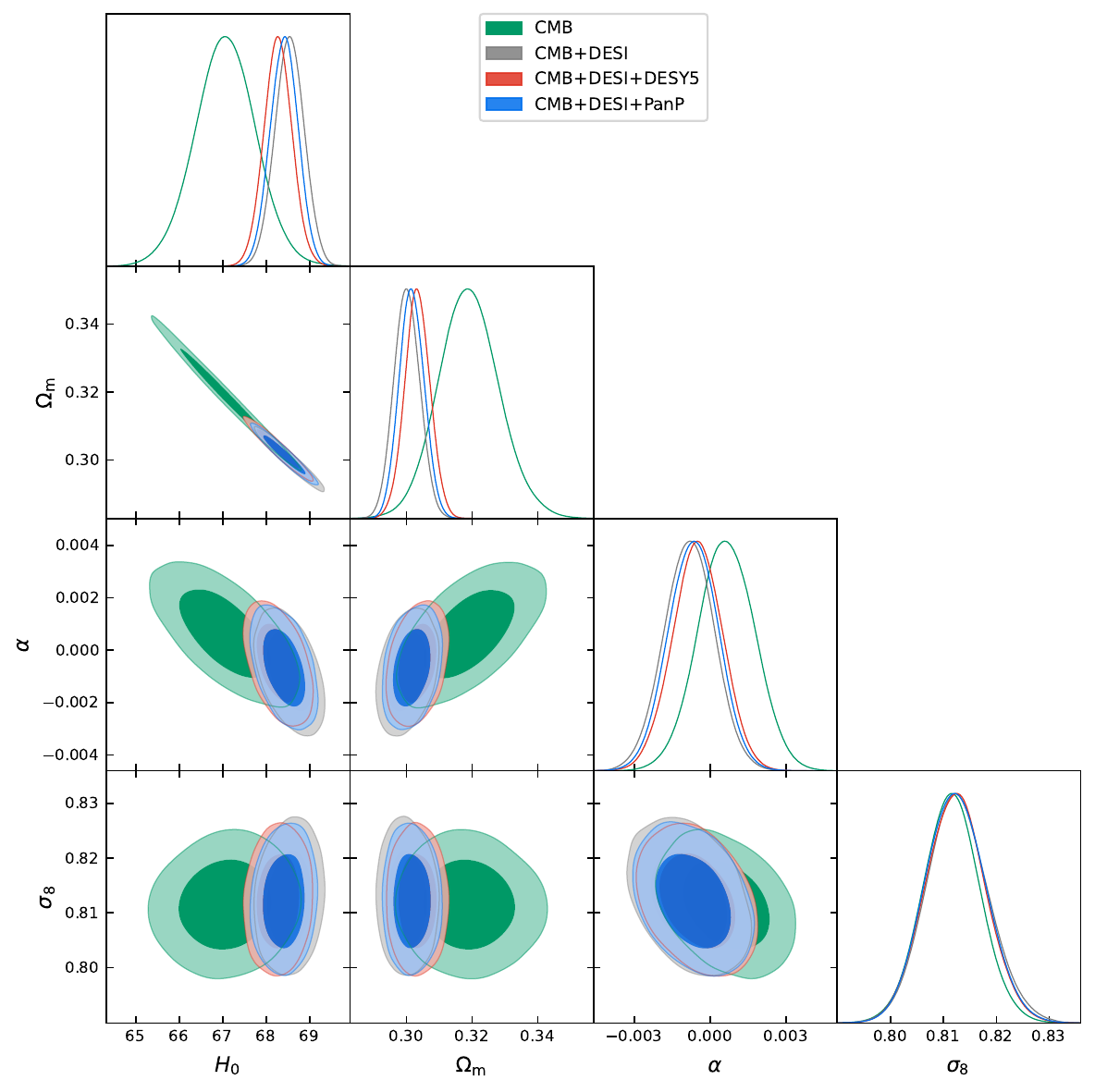}}
\end{center}
\caption{The one and two-dimensional marginalised posterior distributions in a torsion cosmology from the different datasets.}  \label{torsion2}
\end{figure}

Our analysis of the torsion cosmology model, combining the DESI, PantheonPlus/DESY5, and CMB datasets, yields robust constraints on key cosmological parameters. 

For comparison, we first present the constraints on the standard $\Lambda$CDM model under various dataset combinations (Table \ref{tab1}), and the 1D marginalized posterior distributions and 2D joint confidence regions are shown in Fig. \ref{torsion1}. The CMB-alone data yield a Hubble constant of $H_0 = 67.31\pm 0.53~\mathrm{km~s^{-1}~Mpc^{-1}}$, matter density parameter $\Omega_{\mathrm{m}} = 0.3160\pm 0.0073$, and amplitude of matter clustering $\sigma_8 = 0.8127\pm 0.0052$, with a goodness of fit $\chi^2_{\mathrm{min}}/N_{\rm dof} = 1.207$. The addition of DESI BAO data (CMB+DESI) significantly tightens these constraints, resulting in $H_0 = 68.40\pm 0.29$, $\Omega_{\mathrm{m}} = 0.3013\pm 0.0037$, and $\sigma_8 = 0.8110\pm 0.0055$, while the goodness of fit is $\chi^2_{\mathrm{min}}/N_{\rm dof} = 1.208$. Further incorporating the DESY5 supernova data (CMB+DESI+DESY5) gives $H_0 = 68.20\pm 0.29$, $\Omega_{\mathrm{m}} = 0.3038\pm 0.0036$, $\sigma_8 = 0.8112\pm 0.0054$, and $\chi^2_{\mathrm{min}}/N_{\rm dof} = 1.108$. The full combination with PantheonPlus (CMB+DESI+PantheonPlus) yields $H_0 = 68.31\pm 0.29$, $\Omega_{\mathrm{m}} = 0.3025\pm 0.0036$, $\sigma_8 = 0.8110\pm 0.0055$, and $\chi^2_{\mathrm{min}}/N_{\rm dof} = 1.016$. Notably, the reduced $\chi^2$ values ($\chi^2_{\mathrm{min}}/N_{\mathrm{dof}}$) range from 1.016 to 1.208 across different dataset combinations, indicating good fit quality for the $\Lambda$CDM model.

Turning to the torsion cosmology model, the CMB-alone analysis produces weak constraints on the torsion parameter $\alpha = 0.0006\pm 0.0011$ and Hubble constant $H_0 = 67.06\pm 0.70~\mathrm{km~s^{-1}~Mpc^{-1}}$. Large uncertainties reflect the well-known geometric degeneracy inherent in CMB data. Notably, the matter density $\Omega_{\mathrm{m}} = 0.3190\pm 0.0092$ is higher than typical late-time measurements, while 
$\chi^2_{\mathrm{min}}/N_{\mathrm{dof}} = 1.204$ indicates a good quality of fit, demonstrating that CMB data alone can accommodate the torsion model but with substantial parameter freedom. Adding DESI BAO data (CMB+DESI) significantly tightens constraints, reducing the uncertainty of $\alpha$ and improving the precision of the Hubble constant to $H_0 = 68.54\pm 0.33\ \mathrm{km\ s^{-1}\ Mpc^{-1}}$. The matter density parameter decreases to $\Omega_{\mathrm{m}} = 0.3001\pm 0.0039$, aligning with standard values. The combination CMB+DESI+DESY5 further refines the parameters, yielding $\alpha = -0.00050\pm 0.00098$, $H_0 = 68.27\pm 0.32$, and $\Omega_{\mathrm{m}} = 0.3033\pm 0.0039$, with $\chi^2_{\mathrm{min}}/N_{\mathrm{dof}} = 1.106$. Finally, the full dataset combination CMB+DESI+PantheonPlus provides the most robust constraints: $\alpha = -0.00066\pm 0.00098$ (consistent with $\Lambda$CDM at $<1\sigma$), $H_0 = 68.41\pm 0.32\ \mathrm{km\ s^{-1}\ Mpc^{-1}}$, $\sigma_8 = 0.8123\pm 0.0057$, and $\chi^2_{\mathrm{min}}/N_{\mathrm{dof}} = 1.013$.
The 1D marginalized posterior distributions and 2D joint confidence regions are displayed in Fig. \ref{torsion2}, and numerical results of the constraints are shown in Table \ref{tab2}.

The introduction of DESI BAO data (CMB+DESI) acts as a powerful low-redshift anchor, breaking geometric degeneracies through precise measurements of $D_M(z)/r_d$ and $H(z)r_d$ across multiple redshifts. This dramatically tightens parameter constraints, with Hubble constant converging to $H_0 = 68.54\pm 0.33$ km s$^{-1}$ Mpc$^{-1}$ and the torsion parameter shifting to $\alpha = -0.00083\pm 0.00099$. The negative $\alpha$ value amplifies the effective Hubble friction term in the growth equation, potentially enhancing structure formation, though this effect is precisely counterbalanced by a lower matter density $\Omega_{\mathrm{m}} = 0.3001\pm 0.0039$ to maintain consistency with $\Lambda$CDM's $\sigma_8$ predictions. The model's key achievement lies in its simultaneous adjustment of late-time expansion and growth history: $H_0$ converges to $68.41\pm 0.32$ km s$^{-1}$ Mpc$^{-1}$ by subtly modifying the distance to last scattering and the integrated Sachs-Wolfe effect, while predicting $\sigma_8 = 0.8123 \pm 0.0057$ ($S_8 = 0.812\pm 0.006$) that shows exceptional concordance with KiDS-1000 results ($S_8 = 0.815^{+0.016}_{-0.021}$) \citep{2025arXiv250319441W}, reducing the $S_8$ tension with Planck from $\sim 2.3\sigma$ to $0.1\sigma$. This resolution is achieved through the unique scale-independent, time-dependent modulation of the $\alpha H$ term, which suppresses power on nonlinear scales without requiring new particles or early dark energy—a distinctive mechanism among $\Lambda$CDM extensions.

The subsequent additions of DESY5 and PantheonPlus supernova data further refine the cosmic expansion history. Supernovae provide a direct measure of the luminosity distance, which is integrated over the Hubble function $H(z)$. The consistency of the constraints across CMB+DESI+DESY5 and CMB+DESI+PantheonPlus combinations ($\alpha\sim -0.0006$, $H_0\sim 68.4$ km s$^{-1}$ Mpc$^{-1}$, $\Omega_m \sim 0.302$) demonstrates the robustness of the torsion framework to different cosmological probes. The full dataset combination provides the most robust constraints: $\alpha = -0.00066\pm 0.00098$, consistent with zero at less than $1\sigma$, effectively passing a stringent test of its deviation from $\Lambda$CDM.
The findings in the torsion cosmology model reveal several key results. First, the torsion parameter $\alpha$ is consistently constrained to be near zero across all datasets, with its uncertainty drastically reduced by the addition of low-redshift probes. The Hubble constant converges to values around $H_0\sim68.3-68.5~\mathrm{km~s^{-1}~Mpc^{-1}}$, in excellent agreement with Planck $\Lambda$CDM. Second, the matter sector parameters demonstrate full consistency with $\Lambda$CDM expectations. We find $\Omega_\mathrm{m}\approx0.301-0.303$ and $\sigma_8\approx0.812$ for the combined datasets, showing no significant tension with Planck measurements. The primordial parameters remain stable across datasets, with $n_s\approx0.971$ and $\log(10^{10}A_s)\approx3.057$, maintaining agreement with CMB constraints.

The model comparison based on AIC values demonstrates a consistent and statistically significant preference for the torsion model over the $\Lambda$CDM cosmology across all observational datasets. The $\Delta$AIC values, calculated as AIC${\rm torsion}$ - AIC${\rm \Lambda CDM}$, are substantially negative in every case: $-6.16$ for CMB alone, $-5.68$ for CMB+DESI, $-6.04$ for CMB+DESI+DESY5, and $-6.62$ for CMB+DESI+PantheonPlus. According to the Akaike Information Criterion, these negative values indicate that the torsion model provides a better fit to the data after accounting for the additional parameter complexity.
The improvement in AIC is particularly notable given the comprehensive nature of the combined datasets. The CMB+DESI+PantheonPlus combination shows the strongest preference for the torsion model ($\Delta$AIC = $-6.62$), suggesting that the inclusion of both baryon acoustic oscillation and supernova measurements enhances the model's ability to describe cosmological observations. The consistency of negative $\Delta$AIC values across all dataset combinations indicates that the torsion parameter $\alpha$ contributes meaningfully to fitting diverse cosmological probes, from the early universe (CMB) to the late-time expansion history (DESI, PantheonPlus, DESY5).
These results provide compelling evidence that the torsion model represents a statistically superior framework for describing the current cosmological data compared to the standard $\Lambda$CDM model.

\section{Conclusion}\label{sec:5}
Our analysis of the torsion cosmology framework, utilizing the latest CMB data combination (Planck 2018 TT, TE, EE spectra together with the joint NPIPE PR4 + ACT DR6 lensing potential power spectrum), in combination with DESI BAO and PantheonPlus supernovae, constrains the torsion parameter to $\alpha = -0.00066\pm 0.00098$, which remains consistent with zero within $1\sigma$. The value of $\alpha$ shifts from slightly positive when constrained by CMB data alone to slightly negative when including low-redshift probes (DESI and SN Ia), reflecting the breaking of geometric degeneracies and the additional constraining power of these datasets, rather than indicating a redshift-dependent evolution of the parameter itself (which is constant in the model). 

The enhanced constraining power of our combined CMB dataset, particularly through the inclusion of the precise lensing measurements from ACT DR6 and NPIPE PR4, provides a robust anchor for testing the torsion framework. The model demonstrates considerable promise in addressing key cosmological tensions. The Hubble constant converges to $H_0 = 68.41\pm 0.32~\mathrm{km~s^{-1}~Mpc^{-1}}$ with the full dataset combination, effectively bridging early and late-universe measurements. More notably, the framework shows particular strength in alleviating the $S_8$ tension: for the CMB+DESI+PantheonPlus combination, we find $S_8 = \sigma_8\sqrt{\Omega_m/0.3} = 0.812\pm 0.006$, reducing the discrepancy with KiDS-1000 results ($S_8=0.815^{+0.016}_{-0.021}$) from $\sim 2.3\sigma$ in $\Lambda$CDM to just $0.1\sigma$. This resolution is achieved through the unique scale-dependent damping introduced by the torsion term $\alpha H$ in the growth equation, which naturally suppresses structure formation without compromising the fit to CMB measurements—a distinctive advantage over many other $\Lambda$CDM extensions.

The torsion model shows improved fits to all datasets according to the Akaike information criterion, as evidenced by the consistent negative $\Delta$AIC values ranging from -5.68 to -6.62 relative to $\Lambda$CDM. The strongest preference is seen in the CMB+DESI+PantheonPlus combination ($\Delta$AIC = -6.62), while the CMB+DESI dataset shows a slightly less pronounced but still significant improvement ($\Delta$AIC = -5.68). The CMB-only and CMB+DESI+DESY5 combinations also show substantial improvements with $\Delta$AIC values of -6.16 and -6.04 respectively. This consistent pattern across all dataset combinations indicates that the torsion parameter $\alpha$ provides meaningful improvement in fitting diverse cosmological observations, from early universe constraints (CMB) to late-time expansion measurements (DESI, PantheonPlus, DESY5). The results suggest that the torsion model represents a statistically superior framework according to the Akaike criterion, with the most comprehensive dataset combination providing the strongest evidence for its preference over the standard $\Lambda$CDM cosmology.

Future data from ongoing and upcoming surveys (Euclid, Roman, Rubin Observatory) could help resolve this tension. These missions will provide exquisitely precise measurements of the growth of structure across a wide redshift range, offering direct tests of the redshift-dependent behavior predicted by the $\alpha H$ term. Additionally, improved BAO measurements from DESI and future spectrographs will better constrain the expansion history, potentially breaking degeneracies and yielding tighter constraints on $\alpha$. The model's implications for cosmic acceleration merit particular investigation, as they could provide distinguishing features from other dark energy models.
While current results do not conclusively favor torsion cosmology over $\Lambda$CDM, the framework's ability to simultaneously address multiple cosmological tensions while maintaining excellent fit quality makes it a compelling extension worthy of continued investigation. The enhanced precision of our CMB dataset, particularly through the inclusion of the latest lensing measurements, provides a robust foundation for these constraints. The model represents a physically motivated alternative that demonstrates how subtle modifications to gravity's role in structure formation can yield improved concordance across cosmological datasets.
\vspace{0.5cm}
\section{Acknowledgments}
This work was supported by the National Natural Science Foundation of China under Grants No. 12203009, No. 12475051, and No. 12035005; The National Key Research and Development Program of China (No. 2024YFC2207400); The Chutian Scholars Program in Hubei Province (X2023007); The Innovative Research Group of Hunan Province under Grant No. 2024JJ1006; The Science and Technology Innovation Program of Hunan Province under Grant No. 2024RC1050.

\bibliographystyle{apsrev} 
\bibliography{references}  

\begin{thebibliography}{81}
\expandafter\ifx\csname natexlab\endcsname\relax\def\natexlab#1{#1}\fi
\expandafter\ifx\csname bibnamefont\endcsname\relax
  \def\bibnamefont#1{#1}\fi
\expandafter\ifx\csname bibfnamefont\endcsname\relax
  \def\bibfnamefont#1{#1}\fi
\expandafter\ifx\csname citenamefont\endcsname\relax
  \def\citenamefont#1{#1}\fi
\expandafter\ifx\csname url\endcsname\relax
  \def\url#1{\texttt{#1}}\fi
\expandafter\ifx\csname urlprefix\endcsname\relax\def\urlprefix{URL }\fi
\providecommand{\bibinfo}[2]{#2}
\providecommand{\eprint}[2][]{\url{#2}}

\bibitem[{\citenamefont{{Riess} et~al.}(1998)\citenamefont{{Riess}, {Filippenko}, {Challis}, {Clocchiatti}, {Diercks}, {Garnavich}, {Gilliland}, {Hogan}, {Jha}, {Kirshner} et~al.}}]{1998AJ....116.1009R}
\bibinfo{author}{\bibfnamefont{A.~G.} \bibnamefont{{Riess}}}, \bibinfo{author}{\bibfnamefont{A.~V.} \bibnamefont{{Filippenko}}}, \bibinfo{author}{\bibfnamefont{P.}~\bibnamefont{{Challis}}}, \bibinfo{author}{\bibfnamefont{A.}~\bibnamefont{{Clocchiatti}}}, \bibinfo{author}{\bibfnamefont{A.}~\bibnamefont{{Diercks}}}, \bibinfo{author}{\bibfnamefont{P.~M.} \bibnamefont{{Garnavich}}}, \bibinfo{author}{\bibfnamefont{R.~L.} \bibnamefont{{Gilliland}}}, \bibinfo{author}{\bibfnamefont{C.~J.} \bibnamefont{{Hogan}}}, \bibinfo{author}{\bibfnamefont{S.}~\bibnamefont{{Jha}}}, \bibinfo{author}{\bibfnamefont{R.~P.} \bibnamefont{{Kirshner}}}, \bibnamefont{et~al.}, \bibinfo{journal}{\aj} \textbf{\bibinfo{volume}{116}}, \bibinfo{pages}{1009} (\bibinfo{year}{1998}), \eprint{astro-ph/9805201}.

\bibitem[{\citenamefont{{Perlmutter} et~al.}(1999)\citenamefont{{Perlmutter}, {Aldering}, {Goldhaber}, {Knop}, {Nugent}, {Castro}, {Deustua}, {Fabbro}, {Goobar}, {Groom} et~al.}}]{1999ApJ...517..565P}
\bibinfo{author}{\bibfnamefont{S.}~\bibnamefont{{Perlmutter}}}, \bibinfo{author}{\bibfnamefont{G.}~\bibnamefont{{Aldering}}}, \bibinfo{author}{\bibfnamefont{G.}~\bibnamefont{{Goldhaber}}}, \bibinfo{author}{\bibfnamefont{R.~A.} \bibnamefont{{Knop}}}, \bibinfo{author}{\bibfnamefont{P.}~\bibnamefont{{Nugent}}}, \bibinfo{author}{\bibfnamefont{P.~G.} \bibnamefont{{Castro}}}, \bibinfo{author}{\bibfnamefont{S.}~\bibnamefont{{Deustua}}}, \bibinfo{author}{\bibfnamefont{S.}~\bibnamefont{{Fabbro}}}, \bibinfo{author}{\bibfnamefont{A.}~\bibnamefont{{Goobar}}}, \bibinfo{author}{\bibfnamefont{D.~E.} \bibnamefont{{Groom}}}, \bibnamefont{et~al.}, \bibinfo{journal}{\apj} \textbf{\bibinfo{volume}{517}}, \bibinfo{pages}{565} (\bibinfo{year}{1999}), \eprint{astro-ph/9812133}.

\bibitem[{\citenamefont{{Planck Collaboration} et~al.}(2020{\natexlab{a}})\citenamefont{{Planck Collaboration}, {Aghanim}, {Akrami}, {Ashdown}, {Aumont}, {Baccigalupi}, {Ballardini}, {Banday}, {Barreiro}, {Bartolo} et~al.}}]{2020A&A...641A...6P}
\bibinfo{author}{\bibnamefont{{Planck Collaboration}}}, \bibinfo{author}{\bibfnamefont{N.}~\bibnamefont{{Aghanim}}}, \bibinfo{author}{\bibfnamefont{Y.}~\bibnamefont{{Akrami}}}, \bibinfo{author}{\bibfnamefont{M.}~\bibnamefont{{Ashdown}}}, \bibinfo{author}{\bibfnamefont{J.}~\bibnamefont{{Aumont}}}, \bibinfo{author}{\bibfnamefont{C.}~\bibnamefont{{Baccigalupi}}}, \bibinfo{author}{\bibfnamefont{M.}~\bibnamefont{{Ballardini}}}, \bibinfo{author}{\bibfnamefont{A.~J.} \bibnamefont{{Banday}}}, \bibinfo{author}{\bibfnamefont{R.~B.} \bibnamefont{{Barreiro}}}, \bibinfo{author}{\bibfnamefont{N.}~\bibnamefont{{Bartolo}}}, \bibnamefont{et~al.}, \bibinfo{journal}{\aap} \textbf{\bibinfo{volume}{641}}, \bibinfo{eid}{A6} (\bibinfo{year}{2020}{\natexlab{a}}), \eprint{1807.06209}.

\bibitem[{\citenamefont{{Alam} et~al.}(2021)\citenamefont{{Alam}, {Aubert}, {Avila}, {Balland}, {Bautista}, {Bershady}, {Bizyaev}, {Blanton}, {Bolton}, {Bovy} et~al.}}]{2021PhRvD.103h3533A}
\bibinfo{author}{\bibfnamefont{S.}~\bibnamefont{{Alam}}}, \bibinfo{author}{\bibfnamefont{M.}~\bibnamefont{{Aubert}}}, \bibinfo{author}{\bibfnamefont{S.}~\bibnamefont{{Avila}}}, \bibinfo{author}{\bibfnamefont{C.}~\bibnamefont{{Balland}}}, \bibinfo{author}{\bibfnamefont{J.~E.} \bibnamefont{{Bautista}}}, \bibinfo{author}{\bibfnamefont{M.~A.} \bibnamefont{{Bershady}}}, \bibinfo{author}{\bibfnamefont{D.}~\bibnamefont{{Bizyaev}}}, \bibinfo{author}{\bibfnamefont{M.~R.} \bibnamefont{{Blanton}}}, \bibinfo{author}{\bibfnamefont{A.~S.} \bibnamefont{{Bolton}}}, \bibinfo{author}{\bibfnamefont{J.}~\bibnamefont{{Bovy}}}, \bibnamefont{et~al.}, \bibinfo{journal}{\prd} \textbf{\bibinfo{volume}{103}}, \bibinfo{eid}{083533} (\bibinfo{year}{2021}), \eprint{2007.08991}.

\bibitem[{\citenamefont{{Riess} et~al.}(2022)\citenamefont{{Riess}, {Yuan}, {Macri}, {Scolnic}, {Brout}, {Casertano}, {Jones}, {Murakami}, {Anand}, {Breuval} et~al.}}]{2022ApJ...934L...7R}
\bibinfo{author}{\bibfnamefont{A.~G.} \bibnamefont{{Riess}}}, \bibinfo{author}{\bibfnamefont{W.}~\bibnamefont{{Yuan}}}, \bibinfo{author}{\bibfnamefont{L.~M.} \bibnamefont{{Macri}}}, \bibinfo{author}{\bibfnamefont{D.}~\bibnamefont{{Scolnic}}}, \bibinfo{author}{\bibfnamefont{D.}~\bibnamefont{{Brout}}}, \bibinfo{author}{\bibfnamefont{S.}~\bibnamefont{{Casertano}}}, \bibinfo{author}{\bibfnamefont{D.~O.} \bibnamefont{{Jones}}}, \bibinfo{author}{\bibfnamefont{Y.}~\bibnamefont{{Murakami}}}, \bibinfo{author}{\bibfnamefont{G.~S.} \bibnamefont{{Anand}}}, \bibinfo{author}{\bibfnamefont{L.}~\bibnamefont{{Breuval}}}, \bibnamefont{et~al.}, \bibinfo{journal}{\apjl} \textbf{\bibinfo{volume}{934}}, \bibinfo{eid}{L7} (\bibinfo{year}{2022}), \eprint{2112.04510}.

\bibitem[{\citenamefont{{Poulin} et~al.}(2019)\citenamefont{{Poulin}, {Smith}, {Karwal}, and {Kamionkowski}}}]{2019PhRvL.122v1301P}
\bibinfo{author}{\bibfnamefont{V.}~\bibnamefont{{Poulin}}}, \bibinfo{author}{\bibfnamefont{T.~L.} \bibnamefont{{Smith}}}, \bibinfo{author}{\bibfnamefont{T.}~\bibnamefont{{Karwal}}}, \bibnamefont{and} \bibinfo{author}{\bibfnamefont{M.}~\bibnamefont{{Kamionkowski}}}, \bibinfo{journal}{\prl} \textbf{\bibinfo{volume}{122}}, \bibinfo{eid}{221301} (\bibinfo{year}{2019}), \eprint{1811.04083}.

\bibitem[{\citenamefont{{Feeney} et~al.}(2019)\citenamefont{{Feeney}, {Peiris}, {Williamson}, {Nissanke}, {Mortlock}, {Alsing}, and {Scolnic}}}]{2019PhRvL.122f1105F}
\bibinfo{author}{\bibfnamefont{S.~M.} \bibnamefont{{Feeney}}}, \bibinfo{author}{\bibfnamefont{H.~V.} \bibnamefont{{Peiris}}}, \bibinfo{author}{\bibfnamefont{A.~R.} \bibnamefont{{Williamson}}}, \bibinfo{author}{\bibfnamefont{S.~M.} \bibnamefont{{Nissanke}}}, \bibinfo{author}{\bibfnamefont{D.~J.} \bibnamefont{{Mortlock}}}, \bibinfo{author}{\bibfnamefont{J.}~\bibnamefont{{Alsing}}}, \bibnamefont{and} \bibinfo{author}{\bibfnamefont{D.}~\bibnamefont{{Scolnic}}}, \bibinfo{journal}{\prl} \textbf{\bibinfo{volume}{122}}, \bibinfo{eid}{061105} (\bibinfo{year}{2019}), \eprint{1802.03404}.

\bibitem[{\citenamefont{{Dainotti} et~al.}(2023)\citenamefont{{Dainotti}, {Bargiacchi}, {Bogdan}, {Capozziello}, and {Nagataki}}}]{2023arXiv230306974D}
\bibinfo{author}{\bibfnamefont{M.~G.} \bibnamefont{{Dainotti}}}, \bibinfo{author}{\bibfnamefont{G.}~\bibnamefont{{Bargiacchi}}}, \bibinfo{author}{\bibfnamefont{M.}~\bibnamefont{{Bogdan}}}, \bibinfo{author}{\bibfnamefont{S.}~\bibnamefont{{Capozziello}}}, \bibnamefont{and} \bibinfo{author}{\bibfnamefont{S.}~\bibnamefont{{Nagataki}}}, \bibinfo{journal}{arXiv e-prints} \bibinfo{eid}{arXiv:2303.06974} (\bibinfo{year}{2023}), \eprint{2303.06974}.

\bibitem[{\citenamefont{{Liu} et~al.}(2022)\citenamefont{{Liu}, {Cao}, {Biesiada}, and {Geng}}}]{2022ApJ...939...37L}
\bibinfo{author}{\bibfnamefont{T.}~\bibnamefont{{Liu}}}, \bibinfo{author}{\bibfnamefont{S.}~\bibnamefont{{Cao}}}, \bibinfo{author}{\bibfnamefont{M.}~\bibnamefont{{Biesiada}}}, \bibnamefont{and} \bibinfo{author}{\bibfnamefont{S.}~\bibnamefont{{Geng}}}, \bibinfo{journal}{\apj} \textbf{\bibinfo{volume}{939}}, \bibinfo{eid}{37} (\bibinfo{year}{2022}), \eprint{2204.07365}.

\bibitem[{\citenamefont{{Dainotti} et~al.}(2021)\citenamefont{{Dainotti}, {De Simone}, {Schiavone}, {Montani}, {Rinaldi}, and {Lambiase}}}]{2021ApJ...912..150D}
\bibinfo{author}{\bibfnamefont{M.~G.} \bibnamefont{{Dainotti}}}, \bibinfo{author}{\bibfnamefont{B.}~\bibnamefont{{De Simone}}}, \bibinfo{author}{\bibfnamefont{T.}~\bibnamefont{{Schiavone}}}, \bibinfo{author}{\bibfnamefont{G.}~\bibnamefont{{Montani}}}, \bibinfo{author}{\bibfnamefont{E.}~\bibnamefont{{Rinaldi}}}, \bibnamefont{and} \bibinfo{author}{\bibfnamefont{G.}~\bibnamefont{{Lambiase}}}, \bibinfo{journal}{\apj} \textbf{\bibinfo{volume}{912}}, \bibinfo{eid}{150} (\bibinfo{year}{2021}), \eprint{2103.02117}.

\bibitem[{\citenamefont{{Liao} et~al.}(2020)\citenamefont{{Liao}, {Shafieloo}, {Keeley}, and {Linder}}}]{2020ApJ...895L..29L}
\bibinfo{author}{\bibfnamefont{K.}~\bibnamefont{{Liao}}}, \bibinfo{author}{\bibfnamefont{A.}~\bibnamefont{{Shafieloo}}}, \bibinfo{author}{\bibfnamefont{R.~E.} \bibnamefont{{Keeley}}}, \bibnamefont{and} \bibinfo{author}{\bibfnamefont{E.~V.} \bibnamefont{{Linder}}}, \bibinfo{journal}{\apjl} \textbf{\bibinfo{volume}{895}}, \bibinfo{eid}{L29} (\bibinfo{year}{2020}), \eprint{2002.10605}.

\bibitem[{\citenamefont{{Lenart} et~al.}(2023)\citenamefont{{Lenart}, {Bargiacchi}, {Dainotti}, {Nagataki}, and {Capozziello}}}]{2023ApJS..264...46L}
\bibinfo{author}{\bibfnamefont{A.~{\L}.} \bibnamefont{{Lenart}}}, \bibinfo{author}{\bibfnamefont{G.}~\bibnamefont{{Bargiacchi}}}, \bibinfo{author}{\bibfnamefont{M.~G.} \bibnamefont{{Dainotti}}}, \bibinfo{author}{\bibfnamefont{S.}~\bibnamefont{{Nagataki}}}, \bibnamefont{and} \bibinfo{author}{\bibfnamefont{S.}~\bibnamefont{{Capozziello}}}, \bibinfo{journal}{\apjs} \textbf{\bibinfo{volume}{264}}, \bibinfo{eid}{46} (\bibinfo{year}{2023}), \eprint{2211.10785}.

\bibitem[{\citenamefont{{Vagnozzi}}(2020)}]{2020PhRvD.102b3518V}
\bibinfo{author}{\bibfnamefont{S.}~\bibnamefont{{Vagnozzi}}}, \bibinfo{journal}{\prd} \textbf{\bibinfo{volume}{102}}, \bibinfo{eid}{023518} (\bibinfo{year}{2020}), \eprint{1907.07569}.

\bibitem[{\citenamefont{{Vagnozzi}}(2023)}]{2023Univ....9..393V}
\bibinfo{author}{\bibfnamefont{S.}~\bibnamefont{{Vagnozzi}}}, \bibinfo{journal}{Universe} \textbf{\bibinfo{volume}{9}}, \bibinfo{eid}{393} (\bibinfo{year}{2023}), \eprint{2308.16628}.

\bibitem[{\citenamefont{{Banerjee} et~al.}(2021)\citenamefont{{Banerjee}, {Cai}, {Heisenberg}, {Colg{\'a}in}, {Sheikh-Jabbari}, and {Yang}}}]{2021PhRvD.103h1305B}
\bibinfo{author}{\bibfnamefont{A.}~\bibnamefont{{Banerjee}}}, \bibinfo{author}{\bibfnamefont{H.}~\bibnamefont{{Cai}}}, \bibinfo{author}{\bibfnamefont{L.}~\bibnamefont{{Heisenberg}}}, \bibinfo{author}{\bibfnamefont{E.~{\'O}.} \bibnamefont{{Colg{\'a}in}}}, \bibinfo{author}{\bibfnamefont{M.~M.} \bibnamefont{{Sheikh-Jabbari}}}, \bibnamefont{and} \bibinfo{author}{\bibfnamefont{T.}~\bibnamefont{{Yang}}}, \bibinfo{journal}{\prd} \textbf{\bibinfo{volume}{103}}, \bibinfo{eid}{L081305} (\bibinfo{year}{2021}), \eprint{2006.00244}.

\bibitem[{\citenamefont{{Lee} et~al.}(2022)\citenamefont{{Lee}, {Lee}, {{\'O} Colg{\'a}in}, {Sheikh-Jabbari}, and {Thakur}}}]{2022JCAP...04..004L}
\bibinfo{author}{\bibfnamefont{B.-H.} \bibnamefont{{Lee}}}, \bibinfo{author}{\bibfnamefont{W.}~\bibnamefont{{Lee}}}, \bibinfo{author}{\bibfnamefont{E.}~\bibnamefont{{{\'O} Colg{\'a}in}}}, \bibinfo{author}{\bibfnamefont{M.~M.} \bibnamefont{{Sheikh-Jabbari}}}, \bibnamefont{and} \bibinfo{author}{\bibfnamefont{S.}~\bibnamefont{{Thakur}}}, \bibinfo{journal}{\jcap} \textbf{\bibinfo{volume}{2022}}, \bibinfo{eid}{004} (\bibinfo{year}{2022}), \eprint{2202.03906}.

\bibitem[{\citenamefont{{Adame} et~al.}(2025)\citenamefont{{Adame}, {Aguilar}, {Ahlen}, {Alam}, {Alexander}, {Alvarez}, {Alves}, {Anand}, {Andrade}, {Armengaud} et~al.}}]{2025JCAP...02..021A}
\bibinfo{author}{\bibfnamefont{A.~G.} \bibnamefont{{Adame}}}, \bibinfo{author}{\bibfnamefont{J.}~\bibnamefont{{Aguilar}}}, \bibinfo{author}{\bibfnamefont{S.}~\bibnamefont{{Ahlen}}}, \bibinfo{author}{\bibfnamefont{S.}~\bibnamefont{{Alam}}}, \bibinfo{author}{\bibfnamefont{D.~M.} \bibnamefont{{Alexander}}}, \bibinfo{author}{\bibfnamefont{M.}~\bibnamefont{{Alvarez}}}, \bibinfo{author}{\bibfnamefont{O.}~\bibnamefont{{Alves}}}, \bibinfo{author}{\bibfnamefont{A.}~\bibnamefont{{Anand}}}, \bibinfo{author}{\bibfnamefont{U.}~\bibnamefont{{Andrade}}}, \bibinfo{author}{\bibfnamefont{E.}~\bibnamefont{{Armengaud}}}, \bibnamefont{et~al.}, \bibinfo{journal}{\jcap} \textbf{\bibinfo{volume}{2025}}, \bibinfo{eid}{021} (\bibinfo{year}{2025}), \eprint{2404.03002}.

\bibitem[{\citenamefont{{DESI Collaboration} et~al.}(2025)\citenamefont{{DESI Collaboration}, {Abdul-Karim}, {Aguilar}, {Ahlen}, {Alam}, and {Allen}}}]{2025arXiv250314738D}
\bibinfo{author}{\bibnamefont{{DESI Collaboration}}}, \bibinfo{author}{\bibfnamefont{M.}~\bibnamefont{{Abdul-Karim}}}, \bibinfo{author}{\bibfnamefont{J.}~\bibnamefont{{Aguilar}}}, \bibinfo{author}{\bibfnamefont{S.}~\bibnamefont{{Ahlen}}}, \bibinfo{author}{\bibfnamefont{S.}~\bibnamefont{{Alam}}}, \bibnamefont{and} \bibinfo{author}{\bibfnamefont{L.~e.~a.} \bibnamefont{{Allen}}}, \bibinfo{journal}{arXiv e-prints} \bibinfo{eid}{arXiv:2503.14738} (\bibinfo{year}{2025}), \eprint{2503.14738}.

\bibitem[{\citenamefont{{Ye} et~al.}(2025)\citenamefont{{Ye}, {Martinelli}, {Hu}, and {Silvestri}}}]{2025PhRvL.134r1002Y}
\bibinfo{author}{\bibfnamefont{G.}~\bibnamefont{{Ye}}}, \bibinfo{author}{\bibfnamefont{M.}~\bibnamefont{{Martinelli}}}, \bibinfo{author}{\bibfnamefont{B.}~\bibnamefont{{Hu}}}, \bibnamefont{and} \bibinfo{author}{\bibfnamefont{A.}~\bibnamefont{{Silvestri}}}, \bibinfo{journal}{\prl} \textbf{\bibinfo{volume}{134}}, \bibinfo{eid}{181002} (\bibinfo{year}{2025}), \eprint{2407.15832}.

\bibitem[{\citenamefont{{Cai} et~al.}(2025)\citenamefont{{Cai}, {Ren}, {Qiu}, {Li}, and {Zhang}}}]{2025arXiv250524732C}
\bibinfo{author}{\bibfnamefont{Y.}~\bibnamefont{{Cai}}}, \bibinfo{author}{\bibfnamefont{X.}~\bibnamefont{{Ren}}}, \bibinfo{author}{\bibfnamefont{T.}~\bibnamefont{{Qiu}}}, \bibinfo{author}{\bibfnamefont{M.}~\bibnamefont{{Li}}}, \bibnamefont{and} \bibinfo{author}{\bibfnamefont{X.}~\bibnamefont{{Zhang}}}, \bibinfo{journal}{arXiv e-prints} \bibinfo{eid}{arXiv:2505.24732} (\bibinfo{year}{2025}), \eprint{2505.24732}.

\bibitem[{\citenamefont{{Ferrari} et~al.}(2025)\citenamefont{{Ferrari}, {Ballardini}, {Finelli}, and {Paoletti}}}]{2025PhRvD.111h3523F}
\bibinfo{author}{\bibfnamefont{A.~G.} \bibnamefont{{Ferrari}}}, \bibinfo{author}{\bibfnamefont{M.}~\bibnamefont{{Ballardini}}}, \bibinfo{author}{\bibfnamefont{F.}~\bibnamefont{{Finelli}}}, \bibnamefont{and} \bibinfo{author}{\bibfnamefont{D.}~\bibnamefont{{Paoletti}}}, \bibinfo{journal}{\prd} \textbf{\bibinfo{volume}{111}}, \bibinfo{eid}{083523} (\bibinfo{year}{2025}), \eprint{2501.15298}.

\bibitem[{\citenamefont{{Li} et~al.}(2025)\citenamefont{{Li}, {Wang}, {Zhang}, {Saridakis}, and {Cai}}}]{2025arXiv250407791L}
\bibinfo{author}{\bibfnamefont{C.}~\bibnamefont{{Li}}}, \bibinfo{author}{\bibfnamefont{J.}~\bibnamefont{{Wang}}}, \bibinfo{author}{\bibfnamefont{D.}~\bibnamefont{{Zhang}}}, \bibinfo{author}{\bibfnamefont{E.~N.} \bibnamefont{{Saridakis}}}, \bibnamefont{and} \bibinfo{author}{\bibfnamefont{Y.-F.} \bibnamefont{{Cai}}}, \bibinfo{journal}{arXiv e-prints} \bibinfo{eid}{arXiv:2504.07791} (\bibinfo{year}{2025}), \eprint{2504.07791}.

\bibitem[{\citenamefont{{Ishak} et~al.}(2024)\citenamefont{{Ishak}, {Pan}, {Calderon}, {Lodha}, {Valogiannis}, {Aviles}, {Niz}, {Yi}, {Zheng}, {Garcia-Quintero} et~al.}}]{2024arXiv241112026I}
\bibinfo{author}{\bibfnamefont{M.}~\bibnamefont{{Ishak}}}, \bibinfo{author}{\bibfnamefont{J.}~\bibnamefont{{Pan}}}, \bibinfo{author}{\bibfnamefont{R.}~\bibnamefont{{Calderon}}}, \bibinfo{author}{\bibfnamefont{K.}~\bibnamefont{{Lodha}}}, \bibinfo{author}{\bibfnamefont{G.}~\bibnamefont{{Valogiannis}}}, \bibinfo{author}{\bibfnamefont{A.}~\bibnamefont{{Aviles}}}, \bibinfo{author}{\bibfnamefont{G.}~\bibnamefont{{Niz}}}, \bibinfo{author}{\bibfnamefont{L.}~\bibnamefont{{Yi}}}, \bibinfo{author}{\bibfnamefont{C.}~\bibnamefont{{Zheng}}}, \bibinfo{author}{\bibfnamefont{C.}~\bibnamefont{{Garcia-Quintero}}}, \bibnamefont{et~al.}, \bibinfo{journal}{arXiv e-prints} \bibinfo{eid}{arXiv:2411.12026} (\bibinfo{year}{2024}), \eprint{2411.12026}.

\bibitem[{\citenamefont{{Ye}}(2024)}]{2024arXiv241111743Y}
\bibinfo{author}{\bibfnamefont{G.}~\bibnamefont{{Ye}}}, \bibinfo{journal}{arXiv e-prints} \bibinfo{eid}{arXiv:2411.11743} (\bibinfo{year}{2024}), \eprint{2411.11743}.

\bibitem[{\citenamefont{{Giar{\`e}} et~al.}(2024)\citenamefont{{Giar{\`e}}, {Sabogal}, {Nunes}, and {Di Valentino}}}]{2024PhRvL.133y1003G}
\bibinfo{author}{\bibfnamefont{W.}~\bibnamefont{{Giar{\`e}}}}, \bibinfo{author}{\bibfnamefont{M.~A.} \bibnamefont{{Sabogal}}}, \bibinfo{author}{\bibfnamefont{R.~C.} \bibnamefont{{Nunes}}}, \bibnamefont{and} \bibinfo{author}{\bibfnamefont{E.}~\bibnamefont{{Di Valentino}}}, \bibinfo{journal}{\prl} \textbf{\bibinfo{volume}{133}}, \bibinfo{eid}{251003} (\bibinfo{year}{2024}), \eprint{2404.15232}.

\bibitem[{\citenamefont{{Silva} et~al.}(2025{\natexlab{a}})\citenamefont{{Silva}, {Sabogal}, {Scherer}, {Nunes}, {Di Valentino}, and {Kumar}}}]{2025PhRvD.111l3511S}
\bibinfo{author}{\bibfnamefont{E.}~\bibnamefont{{Silva}}}, \bibinfo{author}{\bibfnamefont{M.~A.} \bibnamefont{{Sabogal}}}, \bibinfo{author}{\bibfnamefont{M.}~\bibnamefont{{Scherer}}}, \bibinfo{author}{\bibfnamefont{R.~C.} \bibnamefont{{Nunes}}}, \bibinfo{author}{\bibfnamefont{E.}~\bibnamefont{{Di Valentino}}}, \bibnamefont{and} \bibinfo{author}{\bibfnamefont{S.}~\bibnamefont{{Kumar}}}, \bibinfo{journal}{\prd} \textbf{\bibinfo{volume}{111}}, \bibinfo{eid}{123511} (\bibinfo{year}{2025}{\natexlab{a}}), \eprint{2503.23225}.

\bibitem[{\citenamefont{{Shah} et~al.}(2025)\citenamefont{{Shah}, {Mukherjee}, and {Pal}}}]{2025arXiv250321652S}
\bibinfo{author}{\bibfnamefont{R.}~\bibnamefont{{Shah}}}, \bibinfo{author}{\bibfnamefont{P.}~\bibnamefont{{Mukherjee}}}, \bibnamefont{and} \bibinfo{author}{\bibfnamefont{S.}~\bibnamefont{{Pal}}}, \bibinfo{journal}{arXiv e-prints} \bibinfo{eid}{arXiv:2503.21652} (\bibinfo{year}{2025}), \eprint{2503.21652}.

\bibitem[{\citenamefont{{Li} et~al.}(2024)\citenamefont{{Li}, {Wu}, {Du}, {Jin}, {Li}, {Zhang}, and {Zhang}}}]{2024ApJ...976....1L}
\bibinfo{author}{\bibfnamefont{T.-N.} \bibnamefont{{Li}}}, \bibinfo{author}{\bibfnamefont{P.-J.} \bibnamefont{{Wu}}}, \bibinfo{author}{\bibfnamefont{G.-H.} \bibnamefont{{Du}}}, \bibinfo{author}{\bibfnamefont{S.-J.} \bibnamefont{{Jin}}}, \bibinfo{author}{\bibfnamefont{H.-L.} \bibnamefont{{Li}}}, \bibinfo{author}{\bibfnamefont{J.-F.} \bibnamefont{{Zhang}}}, \bibnamefont{and} \bibinfo{author}{\bibfnamefont{X.}~\bibnamefont{{Zhang}}}, \bibinfo{journal}{\apj} \textbf{\bibinfo{volume}{976}}, \bibinfo{eid}{1} (\bibinfo{year}{2024}), \eprint{2407.14934}.

\bibitem[{\citenamefont{{You} et~al.}(2025)\citenamefont{{You}, {Wang}, and {Yang}}}]{2025arXiv250400985Y}
\bibinfo{author}{\bibfnamefont{C.}~\bibnamefont{{You}}}, \bibinfo{author}{\bibfnamefont{D.}~\bibnamefont{{Wang}}}, \bibnamefont{and} \bibinfo{author}{\bibfnamefont{T.}~\bibnamefont{{Yang}}}, \bibinfo{journal}{arXiv e-prints} \bibinfo{eid}{arXiv:2504.00985} (\bibinfo{year}{2025}), \eprint{2504.00985}.

\bibitem[{\citenamefont{{Pan} and {Ye}}(2025)}]{2025arXiv250319898P}
\bibinfo{author}{\bibfnamefont{J.}~\bibnamefont{{Pan}}} \bibnamefont{and} \bibinfo{author}{\bibfnamefont{G.}~\bibnamefont{{Ye}}}, \bibinfo{journal}{arXiv e-prints} \bibinfo{eid}{arXiv:2503.19898} (\bibinfo{year}{2025}), \eprint{2503.19898}.

\bibitem[{\citenamefont{{Ishiyama} et~al.}(2025)\citenamefont{{Ishiyama}, {Prada}, and {Klypin}}}]{2025arXiv250319352I}
\bibinfo{author}{\bibfnamefont{T.}~\bibnamefont{{Ishiyama}}}, \bibinfo{author}{\bibfnamefont{F.}~\bibnamefont{{Prada}}}, \bibnamefont{and} \bibinfo{author}{\bibfnamefont{A.~A.} \bibnamefont{{Klypin}}}, \bibinfo{journal}{arXiv e-prints} \bibinfo{eid}{arXiv:2503.19352} (\bibinfo{year}{2025}), \eprint{2503.19352}.

\bibitem[{\citenamefont{{Lodha} et~al.}(2025)\citenamefont{{Lodha}, {Calderon}, {Matthewson}, {Shafieloo}, {Ishak}, {Pan}, {Garcia-Quintero}, {Huterer}, {Valogiannis}, {Ure{\~n}a-L{\'o}pez} et~al.}}]{2025arXiv250314743L}
\bibinfo{author}{\bibfnamefont{K.}~\bibnamefont{{Lodha}}}, \bibinfo{author}{\bibfnamefont{R.}~\bibnamefont{{Calderon}}}, \bibinfo{author}{\bibfnamefont{W.~L.} \bibnamefont{{Matthewson}}}, \bibinfo{author}{\bibfnamefont{A.}~\bibnamefont{{Shafieloo}}}, \bibinfo{author}{\bibfnamefont{M.}~\bibnamefont{{Ishak}}}, \bibinfo{author}{\bibfnamefont{J.}~\bibnamefont{{Pan}}}, \bibinfo{author}{\bibfnamefont{C.}~\bibnamefont{{Garcia-Quintero}}}, \bibinfo{author}{\bibfnamefont{D.}~\bibnamefont{{Huterer}}}, \bibinfo{author}{\bibfnamefont{G.}~\bibnamefont{{Valogiannis}}}, \bibinfo{author}{\bibfnamefont{L.~A.} \bibnamefont{{Ure{\~n}a-L{\'o}pez}}}, \bibnamefont{et~al.}, \bibinfo{journal}{arXiv e-prints} \bibinfo{eid}{arXiv:2503.14743} (\bibinfo{year}{2025}), \eprint{2503.14743}.

\bibitem[{\citenamefont{{Liu} et~al.}(2025)\citenamefont{{Liu}, {Li}, and {Wang}}}]{2025arXiv250421373L}
\bibinfo{author}{\bibfnamefont{T.}~\bibnamefont{{Liu}}}, \bibinfo{author}{\bibfnamefont{X.}~\bibnamefont{{Li}}}, \bibnamefont{and} \bibinfo{author}{\bibfnamefont{J.}~\bibnamefont{{Wang}}}, \bibinfo{journal}{arXiv e-prints} \bibinfo{eid}{arXiv:2504.21373} (\bibinfo{year}{2025}), \eprint{2504.21373}.

\bibitem[{\citenamefont{{Cheng} et~al.}(2025)\citenamefont{{Cheng}, {Di Valentino}, {Escamilla}, {Sen}, and {Visinelli}}}]{2025arXiv250502932C}
\bibinfo{author}{\bibfnamefont{H.}~\bibnamefont{{Cheng}}}, \bibinfo{author}{\bibfnamefont{E.}~\bibnamefont{{Di Valentino}}}, \bibinfo{author}{\bibfnamefont{L.~A.} \bibnamefont{{Escamilla}}}, \bibinfo{author}{\bibfnamefont{A.~A.} \bibnamefont{{Sen}}}, \bibnamefont{and} \bibinfo{author}{\bibfnamefont{L.}~\bibnamefont{{Visinelli}}}, \bibinfo{journal}{arXiv e-prints} \bibinfo{eid}{arXiv:2505.02932} (\bibinfo{year}{2025}), \eprint{2505.02932}.

\bibitem[{\citenamefont{{Roy Choudhury} and {Okumura}}(2024)}]{2024ApJ...976L..11R}
\bibinfo{author}{\bibfnamefont{S.}~\bibnamefont{{Roy Choudhury}}} \bibnamefont{and} \bibinfo{author}{\bibfnamefont{T.}~\bibnamefont{{Okumura}}}, \bibinfo{journal}{\apjl} \textbf{\bibinfo{volume}{976}}, \bibinfo{eid}{L11} (\bibinfo{year}{2024}), \eprint{2409.13022}.

\bibitem[{\citenamefont{{Roy Choudhury}}(2025)}]{2025ApJ...986L..31R}
\bibinfo{author}{\bibfnamefont{S.}~\bibnamefont{{Roy Choudhury}}}, \bibinfo{journal}{\apjl} \textbf{\bibinfo{volume}{986}}, \bibinfo{eid}{L31} (\bibinfo{year}{2025}), \eprint{2504.15340}.

\bibitem[{\citenamefont{{Huang} et~al.}(2025)\citenamefont{{Huang}, {Cai}, and {Wang}}}]{2025arXiv250204212H}
\bibinfo{author}{\bibfnamefont{L.}~\bibnamefont{{Huang}}}, \bibinfo{author}{\bibfnamefont{R.-G.} \bibnamefont{{Cai}}}, \bibnamefont{and} \bibinfo{author}{\bibfnamefont{S.-J.} \bibnamefont{{Wang}}}, \bibinfo{journal}{arXiv e-prints} \bibinfo{eid}{arXiv:2502.04212} (\bibinfo{year}{2025}), \eprint{2502.04212}.

\bibitem[{\citenamefont{{Wolf} et~al.}(2025)\citenamefont{{Wolf}, {Garc{\'\i}a-Garc{\'\i}a}, {Anton}, and {Ferreira}}}]{2025arXiv250407679W}
\bibinfo{author}{\bibfnamefont{W.~J.} \bibnamefont{{Wolf}}}, \bibinfo{author}{\bibfnamefont{C.}~\bibnamefont{{Garc{\'\i}a-Garc{\'\i}a}}}, \bibinfo{author}{\bibfnamefont{T.}~\bibnamefont{{Anton}}}, \bibnamefont{and} \bibinfo{author}{\bibfnamefont{P.~G.} \bibnamefont{{Ferreira}}}, \bibinfo{journal}{arXiv e-prints} \bibinfo{eid}{arXiv:2504.07679} (\bibinfo{year}{2025}), \eprint{2504.07679}.

\bibitem[{\citenamefont{{Afroz} and {Mukherjee}}(2025)}]{2025arXiv250416868A}
\bibinfo{author}{\bibfnamefont{S.}~\bibnamefont{{Afroz}}} \bibnamefont{and} \bibinfo{author}{\bibfnamefont{S.}~\bibnamefont{{Mukherjee}}}, \bibinfo{journal}{arXiv e-prints} \bibinfo{eid}{arXiv:2504.16868} (\bibinfo{year}{2025}), \eprint{2504.16868}.

\bibitem[{\citenamefont{{Chakraborty} et~al.}(2025)\citenamefont{{Chakraborty}, {Chanda}, {Das}, and {Dutta}}}]{2025arXiv250310806C}
\bibinfo{author}{\bibfnamefont{A.}~\bibnamefont{{Chakraborty}}}, \bibinfo{author}{\bibfnamefont{P.~K.} \bibnamefont{{Chanda}}}, \bibinfo{author}{\bibfnamefont{S.}~\bibnamefont{{Das}}}, \bibnamefont{and} \bibinfo{author}{\bibfnamefont{K.}~\bibnamefont{{Dutta}}}, \bibinfo{journal}{arXiv e-prints} \bibinfo{eid}{arXiv:2503.10806} (\bibinfo{year}{2025}), \eprint{2503.10806}.

\bibitem[{\citenamefont{{Park} et~al.}(2024{\natexlab{a}})\citenamefont{{Park}, {P{\'e}rez}, and {Ratra}}}]{2024PhRvD.110l3533P}
\bibinfo{author}{\bibfnamefont{C.-G.} \bibnamefont{{Park}}}, \bibinfo{author}{\bibfnamefont{J.~d.~C.} \bibnamefont{{P{\'e}rez}}}, \bibnamefont{and} \bibinfo{author}{\bibfnamefont{B.}~\bibnamefont{{Ratra}}}, \bibinfo{journal}{\prd} \textbf{\bibinfo{volume}{110}}, \bibinfo{eid}{123533} (\bibinfo{year}{2024}{\natexlab{a}}), \eprint{2405.00502}.

\bibitem[{\citenamefont{{Park} et~al.}(2024{\natexlab{b}})\citenamefont{{Park}, {de Cruz Perez}, and {Ratra}}}]{2024arXiv241013627P}
\bibinfo{author}{\bibfnamefont{C.-G.} \bibnamefont{{Park}}}, \bibinfo{author}{\bibfnamefont{J.}~\bibnamefont{{de Cruz Perez}}}, \bibnamefont{and} \bibinfo{author}{\bibfnamefont{B.}~\bibnamefont{{Ratra}}}, \bibinfo{journal}{arXiv e-prints} \bibinfo{eid}{arXiv:2410.13627} (\bibinfo{year}{2024}{\natexlab{b}}), \eprint{2410.13627}.

\bibitem[{\citenamefont{{Park} and {Ratra}}(2025)}]{2025arXiv250103480P}
\bibinfo{author}{\bibfnamefont{C.-G.} \bibnamefont{{Park}}} \bibnamefont{and} \bibinfo{author}{\bibfnamefont{B.}~\bibnamefont{{Ratra}}}, \bibinfo{journal}{arXiv e-prints} \bibinfo{eid}{arXiv:2501.03480} (\bibinfo{year}{2025}), \eprint{2501.03480}.

\bibitem[{\citenamefont{{Yin}}(2022)}]{2022PhRvD.106e5014Y}
\bibinfo{author}{\bibfnamefont{W.}~\bibnamefont{{Yin}}}, \bibinfo{journal}{\prd} \textbf{\bibinfo{volume}{106}}, \bibinfo{eid}{055014} (\bibinfo{year}{2022}), \eprint{2108.04246}.

\bibitem[{\citenamefont{{Yin}}(2024)}]{2024JHEP...05..327Y}
\bibinfo{author}{\bibfnamefont{W.}~\bibnamefont{{Yin}}}, \bibinfo{journal}{Journal of High Energy Physics} \textbf{\bibinfo{volume}{2024}}, \bibinfo{eid}{327} (\bibinfo{year}{2024}), \eprint{2404.06444}.

\bibitem[{\citenamefont{{Lee} et~al.}(2025)\citenamefont{{Lee}, {Murai}, {Takahashi}, and {Yin}}}]{2025arXiv250318417L}
\bibinfo{author}{\bibfnamefont{J.}~\bibnamefont{{Lee}}}, \bibinfo{author}{\bibfnamefont{K.}~\bibnamefont{{Murai}}}, \bibinfo{author}{\bibfnamefont{F.}~\bibnamefont{{Takahashi}}}, \bibnamefont{and} \bibinfo{author}{\bibfnamefont{W.}~\bibnamefont{{Yin}}}, \bibinfo{journal}{arXiv e-prints} \bibinfo{eid}{arXiv:2503.18417} (\bibinfo{year}{2025}), \eprint{2503.18417}.

\bibitem[{\citenamefont{{Jiang} et~al.}(2024)\citenamefont{{Jiang}, {Pedrotti}, {da Costa}, and {Vagnozzi}}}]{2024PhRvD.110l3519J}
\bibinfo{author}{\bibfnamefont{J.-Q.} \bibnamefont{{Jiang}}}, \bibinfo{author}{\bibfnamefont{D.}~\bibnamefont{{Pedrotti}}}, \bibinfo{author}{\bibfnamefont{S.~S.} \bibnamefont{{da Costa}}}, \bibnamefont{and} \bibinfo{author}{\bibfnamefont{S.}~\bibnamefont{{Vagnozzi}}}, \bibinfo{journal}{\prd} \textbf{\bibinfo{volume}{110}}, \bibinfo{eid}{123519} (\bibinfo{year}{2024}), \eprint{2408.02365}.

\bibitem[{\citenamefont{{Jiang} et~al.}(2025)\citenamefont{{Jiang}, {Giar{\`e}}, {Gariazzo}, {Dainotti}, {Di Valentino}, {Mena}, {Pedrotti}, {Santos da Costa}, and {Vagnozzi}}}]{2025JCAP...01..153J}
\bibinfo{author}{\bibfnamefont{J.-Q.} \bibnamefont{{Jiang}}}, \bibinfo{author}{\bibfnamefont{W.}~\bibnamefont{{Giar{\`e}}}}, \bibinfo{author}{\bibfnamefont{S.}~\bibnamefont{{Gariazzo}}}, \bibinfo{author}{\bibfnamefont{M.~G.} \bibnamefont{{Dainotti}}}, \bibinfo{author}{\bibfnamefont{E.}~\bibnamefont{{Di Valentino}}}, \bibinfo{author}{\bibfnamefont{O.}~\bibnamefont{{Mena}}}, \bibinfo{author}{\bibfnamefont{D.}~\bibnamefont{{Pedrotti}}}, \bibinfo{author}{\bibfnamefont{S.}~\bibnamefont{{Santos da Costa}}}, \bibnamefont{and} \bibinfo{author}{\bibfnamefont{S.}~\bibnamefont{{Vagnozzi}}}, \bibinfo{journal}{\jcap} \textbf{\bibinfo{volume}{2025}}, \bibinfo{eid}{153} (\bibinfo{year}{2025}), \eprint{2407.18047}.

\bibitem[{\citenamefont{{Chen} et~al.}(2011)\citenamefont{{Chen}, {Zhu}, {Xu}, and {Alcaniz}}}]{2011PhLB..698..175C}
\bibinfo{author}{\bibfnamefont{Y.}~\bibnamefont{{Chen}}}, \bibinfo{author}{\bibfnamefont{Z.-H.} \bibnamefont{{Zhu}}}, \bibinfo{author}{\bibfnamefont{L.}~\bibnamefont{{Xu}}}, \bibnamefont{and} \bibinfo{author}{\bibfnamefont{J.~S.} \bibnamefont{{Alcaniz}}}, \bibinfo{journal}{Physics Letters B} \textbf{\bibinfo{volume}{698}}, \bibinfo{pages}{175} (\bibinfo{year}{2011}), \eprint{1103.2512}.

\bibitem[{\citenamefont{{Medeiros dos Santos} et~al.}(2025)\citenamefont{{Medeiros dos Santos}, {Morais}, {Pan}, {Yang}, and {Di Valentino}}}]{2025arXiv250404646M}
\bibinfo{author}{\bibfnamefont{F.~B.} \bibnamefont{{Medeiros dos Santos}}}, \bibinfo{author}{\bibfnamefont{J.}~\bibnamefont{{Morais}}}, \bibinfo{author}{\bibfnamefont{S.}~\bibnamefont{{Pan}}}, \bibinfo{author}{\bibfnamefont{W.}~\bibnamefont{{Yang}}}, \bibnamefont{and} \bibinfo{author}{\bibfnamefont{E.}~\bibnamefont{{Di Valentino}}}, \bibinfo{journal}{arXiv e-prints} \bibinfo{eid}{arXiv:2504.04646} (\bibinfo{year}{2025}), \eprint{2504.04646}.

\bibitem[{\citenamefont{{Gialamas} et~al.}(2025)\citenamefont{{Gialamas}, {H{\"u}tsi}, {Raidal}, {Urrutia}, {Vasar}, and {Veerm{\"a}e}}}]{2025arXiv250621542G}
\bibinfo{author}{\bibfnamefont{I.~D.} \bibnamefont{{Gialamas}}}, \bibinfo{author}{\bibfnamefont{G.}~\bibnamefont{{H{\"u}tsi}}}, \bibinfo{author}{\bibfnamefont{M.}~\bibnamefont{{Raidal}}}, \bibinfo{author}{\bibfnamefont{J.}~\bibnamefont{{Urrutia}}}, \bibinfo{author}{\bibfnamefont{M.}~\bibnamefont{{Vasar}}}, \bibnamefont{and} \bibinfo{author}{\bibfnamefont{H.}~\bibnamefont{{Veerm{\"a}e}}}, \bibinfo{journal}{arXiv e-prints} \bibinfo{eid}{arXiv:2506.21542} (\bibinfo{year}{2025}), \eprint{2506.21542}.

\bibitem[{\citenamefont{{Colg{\'a}in} et~al.}(2025)\citenamefont{{Colg{\'a}in}, {Pourojaghi}, {Sheikh-Jabbari}, and {Yin}}}]{2025arXiv250404417C}
\bibinfo{author}{\bibfnamefont{E.~{\'O}.} \bibnamefont{{Colg{\'a}in}}}, \bibinfo{author}{\bibfnamefont{S.}~\bibnamefont{{Pourojaghi}}}, \bibinfo{author}{\bibfnamefont{M.~M.} \bibnamefont{{Sheikh-Jabbari}}}, \bibnamefont{and} \bibinfo{author}{\bibfnamefont{L.}~\bibnamefont{{Yin}}}, \bibinfo{journal}{arXiv e-prints} \bibinfo{eid}{arXiv:2504.04417} (\bibinfo{year}{2025}), \eprint{2504.04417}.

\bibitem[{\citenamefont{{Chaussidon} et~al.}(2025)\citenamefont{{Chaussidon}, {White}, {de Mattia}, {Gsponer}, {Ahlen}, {Bianchi}, {Brooks}, {Claybaugh}, {Cole}, {Cuceu} et~al.}}]{2025arXiv250324343C}
\bibinfo{author}{\bibfnamefont{E.}~\bibnamefont{{Chaussidon}}}, \bibinfo{author}{\bibfnamefont{M.}~\bibnamefont{{White}}}, \bibinfo{author}{\bibfnamefont{A.}~\bibnamefont{{de Mattia}}}, \bibinfo{author}{\bibfnamefont{R.}~\bibnamefont{{Gsponer}}}, \bibinfo{author}{\bibfnamefont{S.}~\bibnamefont{{Ahlen}}}, \bibinfo{author}{\bibfnamefont{D.}~\bibnamefont{{Bianchi}}}, \bibinfo{author}{\bibfnamefont{D.}~\bibnamefont{{Brooks}}}, \bibinfo{author}{\bibfnamefont{T.}~\bibnamefont{{Claybaugh}}}, \bibinfo{author}{\bibfnamefont{S.}~\bibnamefont{{Cole}}}, \bibinfo{author}{\bibfnamefont{A.}~\bibnamefont{{Cuceu}}}, \bibnamefont{et~al.}, \bibinfo{journal}{arXiv e-prints} \bibinfo{eid}{arXiv:2503.24343} (\bibinfo{year}{2025}), \eprint{2503.24343}.

\bibitem[{\citenamefont{{Wali Hossain} and {Maqsood}}(2025)}]{2025arXiv250219274W}
\bibinfo{author}{\bibfnamefont{M.}~\bibnamefont{{Wali Hossain}}} \bibnamefont{and} \bibinfo{author}{\bibfnamefont{A.}~\bibnamefont{{Maqsood}}}, \bibinfo{journal}{arXiv e-prints} \bibinfo{eid}{arXiv:2502.19274} (\bibinfo{year}{2025}), \eprint{2502.19274}.

\bibitem[{\citenamefont{{Pang} et~al.}(2025)\citenamefont{{Pang}, {Zhang}, and {Huang}}}]{2025arXiv250321600P}
\bibinfo{author}{\bibfnamefont{Y.-H.} \bibnamefont{{Pang}}}, \bibinfo{author}{\bibfnamefont{X.}~\bibnamefont{{Zhang}}}, \bibnamefont{and} \bibinfo{author}{\bibfnamefont{Q.-G.} \bibnamefont{{Huang}}}, \bibinfo{journal}{arXiv e-prints} \bibinfo{eid}{arXiv:2503.21600} (\bibinfo{year}{2025}), \eprint{2503.21600}.

\bibitem[{\citenamefont{{Nesseris} et~al.}(2025)\citenamefont{{Nesseris}, {Akrami}, and {Starkman}}}]{2025arXiv250322529N}
\bibinfo{author}{\bibfnamefont{S.}~\bibnamefont{{Nesseris}}}, \bibinfo{author}{\bibfnamefont{Y.}~\bibnamefont{{Akrami}}}, \bibnamefont{and} \bibinfo{author}{\bibfnamefont{G.~D.} \bibnamefont{{Starkman}}}, \bibinfo{journal}{arXiv e-prints} \bibinfo{eid}{arXiv:2503.22529} (\bibinfo{year}{2025}), \eprint{2503.22529}.

\bibitem[{\citenamefont{{Silva} et~al.}(2025{\natexlab{b}})\citenamefont{{Silva}, {Sabogal}, {Souza}, {Nunes}, {Di Valentino}, and {Kumar}}}]{2025arXiv250323225S}
\bibinfo{author}{\bibfnamefont{E.}~\bibnamefont{{Silva}}}, \bibinfo{author}{\bibfnamefont{M.~A.} \bibnamefont{{Sabogal}}}, \bibinfo{author}{\bibfnamefont{M.~S.} \bibnamefont{{Souza}}}, \bibinfo{author}{\bibfnamefont{R.~C.} \bibnamefont{{Nunes}}}, \bibinfo{author}{\bibfnamefont{E.}~\bibnamefont{{Di Valentino}}}, \bibnamefont{and} \bibinfo{author}{\bibfnamefont{S.}~\bibnamefont{{Kumar}}}, \bibinfo{journal}{arXiv e-prints} \bibinfo{eid}{arXiv:2503.23225} (\bibinfo{year}{2025}{\natexlab{b}}), \eprint{2503.23225}.

\bibitem[{\citenamefont{{Hammond}}(2002)}]{2002RPPh...65..599H}
\bibinfo{author}{\bibfnamefont{R.~T.} \bibnamefont{{Hammond}}}, \bibinfo{journal}{Reports on Progress in Physics} \textbf{\bibinfo{volume}{65}}, \bibinfo{pages}{599} (\bibinfo{year}{2002}).

\bibitem[{\citenamefont{{Hehl}}(1973)}]{1973GReGr...4..333H}
\bibinfo{author}{\bibfnamefont{F.~W.} \bibnamefont{{Hehl}}}, \bibinfo{journal}{General Relativity and Gravitation} \textbf{\bibinfo{volume}{4}}, \bibinfo{pages}{333} (\bibinfo{year}{1973}).

\bibitem[{\citenamefont{{Kerlick}}(1975)}]{1975PhDT.......104K}
\bibinfo{author}{\bibfnamefont{G.~D.} \bibnamefont{{Kerlick}}}, Ph.D. thesis, \bibinfo{school}{Princeton University, New Jersey} (\bibinfo{year}{1975}).

\bibitem[{\citenamefont{{Kranas} et~al.}(2019)\citenamefont{{Kranas}, {Tsagas}, {Barrow}, and {Iosifidis}}}]{2019EPJC...79..341K}
\bibinfo{author}{\bibfnamefont{D.}~\bibnamefont{{Kranas}}}, \bibinfo{author}{\bibfnamefont{C.~G.} \bibnamefont{{Tsagas}}}, \bibinfo{author}{\bibfnamefont{J.~D.} \bibnamefont{{Barrow}}}, \bibnamefont{and} \bibinfo{author}{\bibfnamefont{D.}~\bibnamefont{{Iosifidis}}}, \bibinfo{journal}{European Physical Journal C} \textbf{\bibinfo{volume}{79}}, \bibinfo{eid}{341} (\bibinfo{year}{2019}), \eprint{1809.10064}.

\bibitem[{\citenamefont{{Kibble}}(1961)}]{1961JMP.....2..212K}
\bibinfo{author}{\bibfnamefont{T.~W.~B.} \bibnamefont{{Kibble}}}, \bibinfo{journal}{Journal of Mathematical Physics} \textbf{\bibinfo{volume}{2}}, \bibinfo{pages}{212} (\bibinfo{year}{1961}).

\bibitem[{\citenamefont{{Sciama}}(1964)}]{1964RvMP...36..463S}
\bibinfo{author}{\bibfnamefont{D.~W.} \bibnamefont{{Sciama}}}, \bibinfo{journal}{Reviews of Modern Physics} \textbf{\bibinfo{volume}{36}}, \bibinfo{pages}{463} (\bibinfo{year}{1964}).

\bibitem[{\citenamefont{{Scolnic} et~al.}(2022)\citenamefont{{Scolnic}, {Brout}, {Carr}, {Riess}, {Davis}, {Dwomoh}, {Jones}, {Ali}, {Charvu}, {Chen} et~al.}}]{2022ApJ...938..113S}
\bibinfo{author}{\bibfnamefont{D.}~\bibnamefont{{Scolnic}}}, \bibinfo{author}{\bibfnamefont{D.}~\bibnamefont{{Brout}}}, \bibinfo{author}{\bibfnamefont{A.}~\bibnamefont{{Carr}}}, \bibinfo{author}{\bibfnamefont{A.~G.} \bibnamefont{{Riess}}}, \bibinfo{author}{\bibfnamefont{T.~M.} \bibnamefont{{Davis}}}, \bibinfo{author}{\bibfnamefont{A.}~\bibnamefont{{Dwomoh}}}, \bibinfo{author}{\bibfnamefont{D.~O.} \bibnamefont{{Jones}}}, \bibinfo{author}{\bibfnamefont{N.}~\bibnamefont{{Ali}}}, \bibinfo{author}{\bibfnamefont{P.}~\bibnamefont{{Charvu}}}, \bibinfo{author}{\bibfnamefont{R.}~\bibnamefont{{Chen}}}, \bibnamefont{et~al.}, \bibinfo{journal}{\apj} \textbf{\bibinfo{volume}{938}}, \bibinfo{eid}{113} (\bibinfo{year}{2022}), \eprint{2112.03863}.

\bibitem[{\citenamefont{{DES Collaboration} et~al.}(2024)\citenamefont{{DES Collaboration}, {Abbott}, {Acevedo}, {Aguena}, {Alarcon}, {Allam}, {Alves}, {Amon}, {Andrade-Oliveira}, {Annis} et~al.}}]{2024ApJ...973L..14D}
\bibinfo{author}{\bibnamefont{{DES Collaboration}}}, \bibinfo{author}{\bibfnamefont{T.~M.~C.} \bibnamefont{{Abbott}}}, \bibinfo{author}{\bibfnamefont{M.}~\bibnamefont{{Acevedo}}}, \bibinfo{author}{\bibfnamefont{M.}~\bibnamefont{{Aguena}}}, \bibinfo{author}{\bibfnamefont{A.}~\bibnamefont{{Alarcon}}}, \bibinfo{author}{\bibfnamefont{S.}~\bibnamefont{{Allam}}}, \bibinfo{author}{\bibfnamefont{O.}~\bibnamefont{{Alves}}}, \bibinfo{author}{\bibfnamefont{A.}~\bibnamefont{{Amon}}}, \bibinfo{author}{\bibfnamefont{F.}~\bibnamefont{{Andrade-Oliveira}}}, \bibinfo{author}{\bibfnamefont{J.}~\bibnamefont{{Annis}}}, \bibnamefont{et~al.}, \bibinfo{journal}{\apjl} \textbf{\bibinfo{volume}{973}}, \bibinfo{eid}{L14} (\bibinfo{year}{2024}), \eprint{2401.02929}.

\bibitem[{\citenamefont{{Planck Collaboration} et~al.}(2020{\natexlab{b}})\citenamefont{{Planck Collaboration}, {Aghanim}, {Akrami}, {Ashdown}, {Aumont}, {Baccigalupi}, {Ballardini}, {Banday}, {Barreiro}, {Bartolo} et~al.}}]{2020A&A...641A...5P}
\bibinfo{author}{\bibnamefont{{Planck Collaboration}}}, \bibinfo{author}{\bibfnamefont{N.}~\bibnamefont{{Aghanim}}}, \bibinfo{author}{\bibfnamefont{Y.}~\bibnamefont{{Akrami}}}, \bibinfo{author}{\bibfnamefont{M.}~\bibnamefont{{Ashdown}}}, \bibinfo{author}{\bibfnamefont{J.}~\bibnamefont{{Aumont}}}, \bibinfo{author}{\bibfnamefont{C.}~\bibnamefont{{Baccigalupi}}}, \bibinfo{author}{\bibfnamefont{M.}~\bibnamefont{{Ballardini}}}, \bibinfo{author}{\bibfnamefont{A.~J.} \bibnamefont{{Banday}}}, \bibinfo{author}{\bibfnamefont{R.~B.} \bibnamefont{{Barreiro}}}, \bibinfo{author}{\bibfnamefont{N.}~\bibnamefont{{Bartolo}}}, \bibnamefont{et~al.}, \bibinfo{journal}{\aap} \textbf{\bibinfo{volume}{641}}, \bibinfo{eid}{A5} (\bibinfo{year}{2020}{\natexlab{b}}), \eprint{1907.12875}.

\bibitem[{\citenamefont{{Planck Collaboration} et~al.}(2020{\natexlab{c}})\citenamefont{{Planck Collaboration}, {Aghanim}, {Akrami}, {Arroja}, {Ashdown}, {Aumont}, {Baccigalupi}, {Ballardini}, {Banday}, {Barreiro} et~al.}}]{2020A&A...641A...1P}
\bibinfo{author}{\bibnamefont{{Planck Collaboration}}}, \bibinfo{author}{\bibfnamefont{N.}~\bibnamefont{{Aghanim}}}, \bibinfo{author}{\bibfnamefont{Y.}~\bibnamefont{{Akrami}}}, \bibinfo{author}{\bibfnamefont{F.}~\bibnamefont{{Arroja}}}, \bibinfo{author}{\bibfnamefont{M.}~\bibnamefont{{Ashdown}}}, \bibinfo{author}{\bibfnamefont{J.}~\bibnamefont{{Aumont}}}, \bibinfo{author}{\bibfnamefont{C.}~\bibnamefont{{Baccigalupi}}}, \bibinfo{author}{\bibfnamefont{M.}~\bibnamefont{{Ballardini}}}, \bibinfo{author}{\bibfnamefont{A.~J.} \bibnamefont{{Banday}}}, \bibinfo{author}{\bibfnamefont{R.~B.} \bibnamefont{{Barreiro}}}, \bibnamefont{et~al.}, \bibinfo{journal}{\aap} \textbf{\bibinfo{volume}{641}}, \bibinfo{eid}{A1} (\bibinfo{year}{2020}{\natexlab{c}}), \eprint{1807.06205}.

\bibitem[{\citenamefont{{Planck Collaboration} et~al.}(2014)\citenamefont{{Planck Collaboration}, {Ade}, {Aghanim}, {Armitage-Caplan}, {Arnaud}, {Ashdown}, {Atrio-Barandela}, {Aumont}, {Baccigalupi}, {Banday} et~al.}}]{Planck2013lensing}
\bibinfo{author}{\bibnamefont{{Planck Collaboration}}}, \bibinfo{author}{\bibfnamefont{P.~A.~R.} \bibnamefont{{Ade}}}, \bibinfo{author}{\bibfnamefont{N.}~\bibnamefont{{Aghanim}}}, \bibinfo{author}{\bibfnamefont{C.}~\bibnamefont{{Armitage-Caplan}}}, \bibinfo{author}{\bibfnamefont{M.}~\bibnamefont{{Arnaud}}}, \bibinfo{author}{\bibfnamefont{M.}~\bibnamefont{{Ashdown}}}, \bibinfo{author}{\bibfnamefont{F.}~\bibnamefont{{Atrio-Barandela}}}, \bibinfo{author}{\bibfnamefont{J.}~\bibnamefont{{Aumont}}}, \bibinfo{author}{\bibfnamefont{C.}~\bibnamefont{{Baccigalupi}}}, \bibinfo{author}{\bibfnamefont{A.~J.} \bibnamefont{{Banday}}}, \bibnamefont{et~al.}, \bibinfo{journal}{\aap} \textbf{\bibinfo{volume}{571}}, \bibinfo{eid}{A17} (\bibinfo{year}{2014}), \eprint{1303.5077}.

\bibitem[{\citenamefont{{Planck Collaboration} et~al.}(2016)\citenamefont{{Planck Collaboration}, {Ade}, {Aghanim}, {Arnaud}, {Ashdown}, {Aumont}, {Baccigalupi}, {Banday}, {Barreiro}, {Bartlett} et~al.}}]{Planck2015lensing}
\bibinfo{author}{\bibnamefont{{Planck Collaboration}}}, \bibinfo{author}{\bibfnamefont{P.~A.~R.} \bibnamefont{{Ade}}}, \bibinfo{author}{\bibfnamefont{N.}~\bibnamefont{{Aghanim}}}, \bibinfo{author}{\bibfnamefont{M.}~\bibnamefont{{Arnaud}}}, \bibinfo{author}{\bibfnamefont{M.}~\bibnamefont{{Ashdown}}}, \bibinfo{author}{\bibfnamefont{J.}~\bibnamefont{{Aumont}}}, \bibinfo{author}{\bibfnamefont{C.}~\bibnamefont{{Baccigalupi}}}, \bibinfo{author}{\bibfnamefont{A.~J.} \bibnamefont{{Banday}}}, \bibinfo{author}{\bibfnamefont{R.~B.} \bibnamefont{{Barreiro}}}, \bibinfo{author}{\bibfnamefont{J.~G.} \bibnamefont{{Bartlett}}}, \bibnamefont{et~al.}, \bibinfo{journal}{\aap} \textbf{\bibinfo{volume}{594}}, \bibinfo{eid}{A15} (\bibinfo{year}{2016}), \eprint{1502.01591}.

\bibitem[{\citenamefont{{Planck Collaboration} et~al.}(2020{\natexlab{d}})\citenamefont{{Planck Collaboration}, {Aghanim}, {Akrami}, {Ashdown}, {Aumont}, {Baccigalupi}, {Ballardini}, {Banday}, {Barreiro}, {Bartolo} et~al.}}]{Planck2018lensing}
\bibinfo{author}{\bibnamefont{{Planck Collaboration}}}, \bibinfo{author}{\bibfnamefont{N.}~\bibnamefont{{Aghanim}}}, \bibinfo{author}{\bibfnamefont{Y.}~\bibnamefont{{Akrami}}}, \bibinfo{author}{\bibfnamefont{M.}~\bibnamefont{{Ashdown}}}, \bibinfo{author}{\bibfnamefont{J.}~\bibnamefont{{Aumont}}}, \bibinfo{author}{\bibfnamefont{C.}~\bibnamefont{{Baccigalupi}}}, \bibinfo{author}{\bibfnamefont{M.}~\bibnamefont{{Ballardini}}}, \bibinfo{author}{\bibfnamefont{A.~J.} \bibnamefont{{Banday}}}, \bibinfo{author}{\bibfnamefont{R.~B.} \bibnamefont{{Barreiro}}}, \bibinfo{author}{\bibfnamefont{N.}~\bibnamefont{{Bartolo}}}, \bibnamefont{et~al.}, \bibinfo{journal}{\aap} \textbf{\bibinfo{volume}{641}}, \bibinfo{eid}{A8} (\bibinfo{year}{2020}{\natexlab{d}}), \eprint{1807.06210}.

\bibitem[{\citenamefont{{Carron} et~al.}(2022)\citenamefont{{Carron}, {Mirmelstein}, and {Lewis}}}]{PlanckPR4}
\bibinfo{author}{\bibfnamefont{J.}~\bibnamefont{{Carron}}}, \bibinfo{author}{\bibfnamefont{M.}~\bibnamefont{{Mirmelstein}}}, \bibnamefont{and} \bibinfo{author}{\bibfnamefont{A.}~\bibnamefont{{Lewis}}}, \bibinfo{journal}{\jcap} \textbf{\bibinfo{volume}{2022}}, \bibinfo{eid}{039} (\bibinfo{year}{2022}), \eprint{2206.07773}.

\bibitem[{\citenamefont{{Madhavacheril} et~al.}(2024)\citenamefont{{Madhavacheril}, {Qu}, {Sherwin}, {MacCrann}, {Li}, {Abril-Cabezas}, {Ade}, {Aiola}, {Alford}, {Amiri} et~al.}}]{ACTDR6lensing1}
\bibinfo{author}{\bibfnamefont{M.~S.} \bibnamefont{{Madhavacheril}}}, \bibinfo{author}{\bibfnamefont{F.~J.} \bibnamefont{{Qu}}}, \bibinfo{author}{\bibfnamefont{B.~D.} \bibnamefont{{Sherwin}}}, \bibinfo{author}{\bibfnamefont{N.}~\bibnamefont{{MacCrann}}}, \bibinfo{author}{\bibfnamefont{Y.}~\bibnamefont{{Li}}}, \bibinfo{author}{\bibfnamefont{I.}~\bibnamefont{{Abril-Cabezas}}}, \bibinfo{author}{\bibfnamefont{P.~A.~R.} \bibnamefont{{Ade}}}, \bibinfo{author}{\bibfnamefont{S.}~\bibnamefont{{Aiola}}}, \bibinfo{author}{\bibfnamefont{T.}~\bibnamefont{{Alford}}}, \bibinfo{author}{\bibfnamefont{M.}~\bibnamefont{{Amiri}}}, \bibnamefont{et~al.}, \bibinfo{journal}{\apj} \textbf{\bibinfo{volume}{962}}, \bibinfo{eid}{113} (\bibinfo{year}{2024}), \eprint{2304.05203}.

\bibitem[{\citenamefont{{Qu} et~al.}(2024)\citenamefont{{Qu}, {Sherwin}, {Madhavacheril}, {Han}, {Crowley}, {Abril-Cabezas}, {Ade}, {Aiola}, {Alford}, {Amiri} et~al.}}]{ACTDR6lensing2}
\bibinfo{author}{\bibfnamefont{F.~J.} \bibnamefont{{Qu}}}, \bibinfo{author}{\bibfnamefont{B.~D.} \bibnamefont{{Sherwin}}}, \bibinfo{author}{\bibfnamefont{M.~S.} \bibnamefont{{Madhavacheril}}}, \bibinfo{author}{\bibfnamefont{D.}~\bibnamefont{{Han}}}, \bibinfo{author}{\bibfnamefont{K.~T.} \bibnamefont{{Crowley}}}, \bibinfo{author}{\bibfnamefont{I.}~\bibnamefont{{Abril-Cabezas}}}, \bibinfo{author}{\bibfnamefont{P.~A.~R.} \bibnamefont{{Ade}}}, \bibinfo{author}{\bibfnamefont{S.}~\bibnamefont{{Aiola}}}, \bibinfo{author}{\bibfnamefont{T.}~\bibnamefont{{Alford}}}, \bibinfo{author}{\bibfnamefont{M.}~\bibnamefont{{Amiri}}}, \bibnamefont{et~al.}, \bibinfo{journal}{\apj} \textbf{\bibinfo{volume}{962}}, \bibinfo{eid}{112} (\bibinfo{year}{2024}), \eprint{2304.05202}.

\bibitem[{\citenamefont{{MacCrann} et~al.}(2024)\citenamefont{{MacCrann}, {Sherwin}, {Qu}, {Namikawa}, {Madhavacheril}, {Abril-Cabezas}, {An}, {Austermann}, {Battaglia}, {Battistelli} et~al.}}]{ACTDR6lensing3}
\bibinfo{author}{\bibfnamefont{N.}~\bibnamefont{{MacCrann}}}, \bibinfo{author}{\bibfnamefont{B.~D.} \bibnamefont{{Sherwin}}}, \bibinfo{author}{\bibfnamefont{F.~J.} \bibnamefont{{Qu}}}, \bibinfo{author}{\bibfnamefont{T.}~\bibnamefont{{Namikawa}}}, \bibinfo{author}{\bibfnamefont{M.~S.} \bibnamefont{{Madhavacheril}}}, \bibinfo{author}{\bibfnamefont{I.}~\bibnamefont{{Abril-Cabezas}}}, \bibinfo{author}{\bibfnamefont{R.}~\bibnamefont{{An}}}, \bibinfo{author}{\bibfnamefont{J.~E.} \bibnamefont{{Austermann}}}, \bibinfo{author}{\bibfnamefont{N.}~\bibnamefont{{Battaglia}}}, \bibinfo{author}{\bibfnamefont{E.~S.} \bibnamefont{{Battistelli}}}, \bibnamefont{et~al.}, \bibinfo{journal}{\apj} \textbf{\bibinfo{volume}{966}}, \bibinfo{eid}{138} (\bibinfo{year}{2024}), \eprint{2304.05196}.

\bibitem[{\citenamefont{{Blas} et~al.}(2011)\citenamefont{{Blas}, {Lesgourgues}, and {Tram}}}]{2011JCAP...07..034B}
\bibinfo{author}{\bibfnamefont{D.}~\bibnamefont{{Blas}}}, \bibinfo{author}{\bibfnamefont{J.}~\bibnamefont{{Lesgourgues}}}, \bibnamefont{and} \bibinfo{author}{\bibfnamefont{T.}~\bibnamefont{{Tram}}}, \bibinfo{journal}{\jcap} \textbf{\bibinfo{volume}{2011}}, \bibinfo{eid}{034} (\bibinfo{year}{2011}), \eprint{1104.2933}.

\bibitem[{\citenamefont{{Torrado} and {Lewis}}(2021)}]{2021JCAP...05..057T}
\bibinfo{author}{\bibfnamefont{J.}~\bibnamefont{{Torrado}}} \bibnamefont{and} \bibinfo{author}{\bibfnamefont{A.}~\bibnamefont{{Lewis}}}, \bibinfo{journal}{\jcap} \textbf{\bibinfo{volume}{2021}}, \bibinfo{eid}{057} (\bibinfo{year}{2021}), \eprint{2005.05290}.

\bibitem[{\citenamefont{{Gelman} and {Rubin}}(1992)}]{1992StaSc...7..457G}
\bibinfo{author}{\bibfnamefont{A.}~\bibnamefont{{Gelman}}} \bibnamefont{and} \bibinfo{author}{\bibfnamefont{D.~B.} \bibnamefont{{Rubin}}}, \bibinfo{journal}{Statistical Science} \textbf{\bibinfo{volume}{7}}, \bibinfo{pages}{457} (\bibinfo{year}{1992}).

\bibitem[{\citenamefont{{Biesiada}}(2007)}]{2007JCAP...02..003B}
\bibinfo{author}{\bibfnamefont{M.}~\bibnamefont{{Biesiada}}}, \bibinfo{journal}{\jcap} \textbf{\bibinfo{volume}{2007}}, \bibinfo{eid}{003} (\bibinfo{year}{2007}), \eprint{astro-ph/0701721}.

\bibitem[{\citenamefont{{Di Valentino} et~al.}(2025)\citenamefont{{Di Valentino}, {Said}, {Riess}, {Pollo}, {Poulin}, {G{\'o}mez-Valent}, {Weltman}, {Palmese}, {Huang}, {Bruck} et~al.}}]{2025PDU....4901965D}
\bibinfo{author}{\bibfnamefont{E.}~\bibnamefont{{Di Valentino}}}, \bibinfo{author}{\bibfnamefont{J.~L.} \bibnamefont{{Said}}}, \bibinfo{author}{\bibfnamefont{A.}~\bibnamefont{{Riess}}}, \bibinfo{author}{\bibfnamefont{A.}~\bibnamefont{{Pollo}}}, \bibinfo{author}{\bibfnamefont{V.}~\bibnamefont{{Poulin}}}, \bibinfo{author}{\bibfnamefont{A.}~\bibnamefont{{G{\'o}mez-Valent}}}, \bibinfo{author}{\bibfnamefont{A.}~\bibnamefont{{Weltman}}}, \bibinfo{author}{\bibfnamefont{A.}~\bibnamefont{{Palmese}}}, \bibinfo{author}{\bibfnamefont{C.~D.} \bibnamefont{{Huang}}}, \bibinfo{author}{\bibfnamefont{C.~v.~d.} \bibnamefont{{Bruck}}}, \bibnamefont{et~al.}, \bibinfo{journal}{Physics of the Dark Universe} \textbf{\bibinfo{volume}{49}}, \bibinfo{eid}{101965} (\bibinfo{year}{2025}), \eprint{2504.01669}.

\bibitem[{\citenamefont{{Schwarz}}(1978)}]{1978AnSta...6..461S}
\bibinfo{author}{\bibfnamefont{G.}~\bibnamefont{{Schwarz}}}, \bibinfo{journal}{Annals of Statistics} \textbf{\bibinfo{volume}{6}}, \bibinfo{pages}{461} (\bibinfo{year}{1978}).

\bibitem[{\citenamefont{{Wright} et~al.}(2025)\citenamefont{{Wright}, {St{\"o}lzner}, {Asgari}, {Bilicki}, {Giblin}, {Heymans}, {Hildebrandt}, {Hoekstra}, {Joachimi}, {Kuijken} et~al.}}]{2025arXiv250319441W}
\bibinfo{author}{\bibfnamefont{A.~H.} \bibnamefont{{Wright}}}, \bibinfo{author}{\bibfnamefont{B.}~\bibnamefont{{St{\"o}lzner}}}, \bibinfo{author}{\bibfnamefont{M.}~\bibnamefont{{Asgari}}}, \bibinfo{author}{\bibfnamefont{M.}~\bibnamefont{{Bilicki}}}, \bibinfo{author}{\bibfnamefont{B.}~\bibnamefont{{Giblin}}}, \bibinfo{author}{\bibfnamefont{C.}~\bibnamefont{{Heymans}}}, \bibinfo{author}{\bibfnamefont{H.}~\bibnamefont{{Hildebrandt}}}, \bibinfo{author}{\bibfnamefont{H.}~\bibnamefont{{Hoekstra}}}, \bibinfo{author}{\bibfnamefont{B.}~\bibnamefont{{Joachimi}}}, \bibinfo{author}{\bibfnamefont{K.}~\bibnamefont{{Kuijken}}}, \bibnamefont{et~al.}, \bibinfo{journal}{arXiv e-prints} \bibinfo{eid}{arXiv:2503.19441} (\bibinfo{year}{2025}), \eprint{2503.19441}.

\end{thebibliography}

\end{document}